\documentclass{emulateapj}
\usepackage{graphicx}
\usepackage{amssymb,amsfonts,amsmath,amstext,amsgen,amsopn,amsxtra,indentfirst,times}

\begin{document}

\title{Carbon Dioxide in Exoplanetary Atmospheres: Rarely Dominant Compared to Carbon Monoxide and Water in Hot, Hydrogen-dominated Atmospheres}

\author{Kevin Heng\altaffilmark{1}}
\author{James R. Lyons\altaffilmark{2}}
\altaffiltext{1}{University of Bern, Center for Space and Habitability, Sidlerstrasse 5, CH-3012, Bern, Switzerland.  Email: kevin.heng@csh.unibe.ch}
\altaffiltext{2}{Arizona State University, School of Earth and Space Exploration, Bateman Physical Sciences, Tempe, AZ 85287-1404, U.S.A.  Email: jimlyons@asu.edu}

\begin{abstract}
We present a comprehensive study of the abundance of carbon dioxide in exoplanetary atmospheres in hot, hydrogen-dominated atmospheres.  We construct novel analytical models of systems in chemical equilibrium that include carbon monoxide, carbon dioxide, water, methane and acetylene and relate the equilibrium constants of the chemical reactions to temperature and pressure via the tabulated Gibbs free energies.  We prove that such chemical systems may be described by a quintic equation for the mixing ratio of methane.  By examining the abundances of these molecules across a broad range of temperatures (spanning equilibrium temperatures from 600 to 2500 K), pressures (via temperature-pressure profiles that explore albedo and opacity variations) and carbon-to-oxygen ratios, we conclude that carbon dioxide is subdominant compared to carbon monoxide and water.  Atmospheric mixing does not alter this conclusion if carbon dioxide is subdominant everywhere in the atmosphere.  Carbon dioxide and carbon monoxide may attain comparable abundances if the metallicity is greatly enhanced, but this property is negated by temperatures above 1000 K.  For hydrogen-dominated atmospheres, our generic result has the implication that retrieval studies may wish to set the subdominance of carbon dioxide as a prior of the calculation and not let its abundance completely roam free as a fitting parameter, because it directly affects the inferred value of the carbon-to-oxygen ratio and may produce unphysical conclusions.  We discuss the relevance of these implications for the hot Jupiter WASP-12b and suggest that some of the previous results are chemically impossible.  The relative abundance of carbon dioxide to acetylene is potentially a sensitive diagnostic of the carbon-to-oxygen ratio.
\end{abstract}


\section{Introduction}

\subsection{Towards robust interpretations of spectra}

There is an ongoing debate in the astrophysical community on how to interpret the measured spectra of the atmospheres of exoplanets.  The first school of thought uses a series of ``forward models": given a set of assumptions about the atmosphere, one computes forward and predicts a synthetic spectrum (e.g., \citealt{ss00,sudarsky00,burrows07,cahoy10,fortney10,barman11,marley12,sb12}).  This approach has the advantage that it is grounded by the laws of physics and chemistry (radiative and chemical equilibria).  It draws from a rich heritage of, and has enjoyed success in, interpreting brown dwarf spectra (e.g., \citealt{marley96,tsuji96,burrows97}).

The disadvantage is that Nature may outsmart our preconceived notions of an atmosphere.  The second school of thought uses ``atmospheric retrieval", which is the attempt to shed these preconceived notions (e.g., \citealt{ms09,bs12,lhi13,line14}).  This approach is sound when remote sensing and in-situ measurements are available, such as for the Earth and the Solar System bodies, but its robustness is not assured when scrutinizing distant, unresolved point sources of light.  For example, workers enforce global energy conservation (e.g., \citealt{ms09}), but this does not guarantee \textit{local} energy conservation (radiative equilibrium\footnote{Whether radiative equilibrium is a good assumption, in three dimensions, is another matter altogether.  But in the context of one-dimensional models, this is a valid discussion to have.}; see Appendix).  Generally, the robustness of an interpretation via retrieval depends on the prior knowledge that one is assuming about the atmosphere---one's model is only as good as the assumptions one inserts into it.

In the present study, our goal is to study the gaseous chemistry of hot, hydrogen-dominated atmospheres.  It may not apply to atmospheres with rocky surfaces, where outgassing may produce carbon-dioxide-dominated atmospheres.  By ``hot", we mean that water cannot condense out, at least in the photospheric regions observed by astronomers.  Furthermore, if the inert gas of the atmosphere is a mixture of hydrogen and helium, we assume that atmospheric escape does not act to remove hydrogen, leaving behind a helium-dominated atmosphere that may contain more carbon dioxide than water \citep{hu14,hu15}.

\subsection{May we set generic priors on carbon dioxide?}
\label{subsect:intro}

To move forward, it would be useful to uncover generic physical and chemical trends that one may insert into these inversion techniques as priors.  In the current paper, we focus on the study of carbon dioxide (CO$_2$), which is typically a minor carrier of carbon in exoplanetary atmospheres.  Understanding its abundance relative to other molecules such as carbon monoxide (CO), water (H$_2$O) and methane (CH$_4$) is directly relevant to deciphering the carbon-to-oxygen ratio (C/O) of an atmosphere \citep{line13},
\begin{equation}
\mbox{C/O} \approx \frac{\tilde{n}_{\rm CH_4} + \tilde{n}_{\rm CO} + \tilde{n}_{\rm CO_2}}{\tilde{n}_{\rm H_2O} + \tilde{n}_{\rm CO} + 2 \tilde{n}_{\rm CO_2}},
\end{equation}
where the mixing ratios are represented by $\tilde{n}$ and labeled with self-explanatory subscripts.  Specifically, if CO is the dominant molecule, then we have $\mbox{C/O}\approx 1$.  By contrast, if CO$_2$ dominates, then we have $\mbox{C/O}\approx 0.5$.  For example, this has direct consequences for the debate on whether WASP-12b is carbon-rich \citep{madhu11} or carbon-poor \citep{line14}; \cite{madhu11} report CO being dominant over CO$_2$ (see their Figure 2), while \cite{line14} report CO$_2$ being dominant over CO (see their Table 3).

\begin{figure}%[!h]
\begin{center}
%\vspace{-0.2in}
\includegraphics[width=\columnwidth]{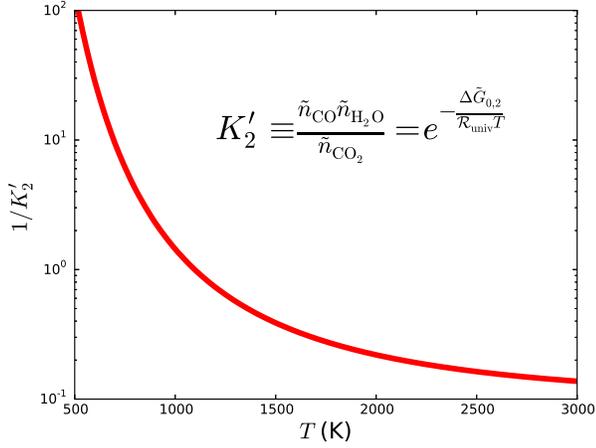}
\end{center}
%\vspace{-0.2in}
\caption{Reciprocal of normalized equilibrium coefficient associated with the production and destruction of carbon dioxide.}
%\vspace{0.1in}
\label{fig:k2}
\end{figure}

\begin{figure}%[!h]
\begin{center}
%\vspace{-0.2in}
\includegraphics[width=\columnwidth]{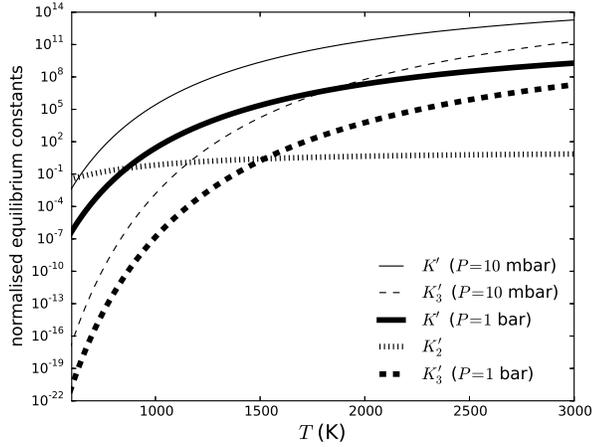}
\end{center}
%\vspace{-0.2in}
\caption{normalized equilibrium constants as a function of temperature for two different values of pressure ($P=10$ mbar and 1 bar).  Note that $K^\prime_2$ has no pressure dependence.}
%\vspace{0.1in}
\label{fig:keq}
\end{figure}

\begin{figure}[!ht]
\begin{center}
%\vspace{-0.2in}
\includegraphics[width=\columnwidth]{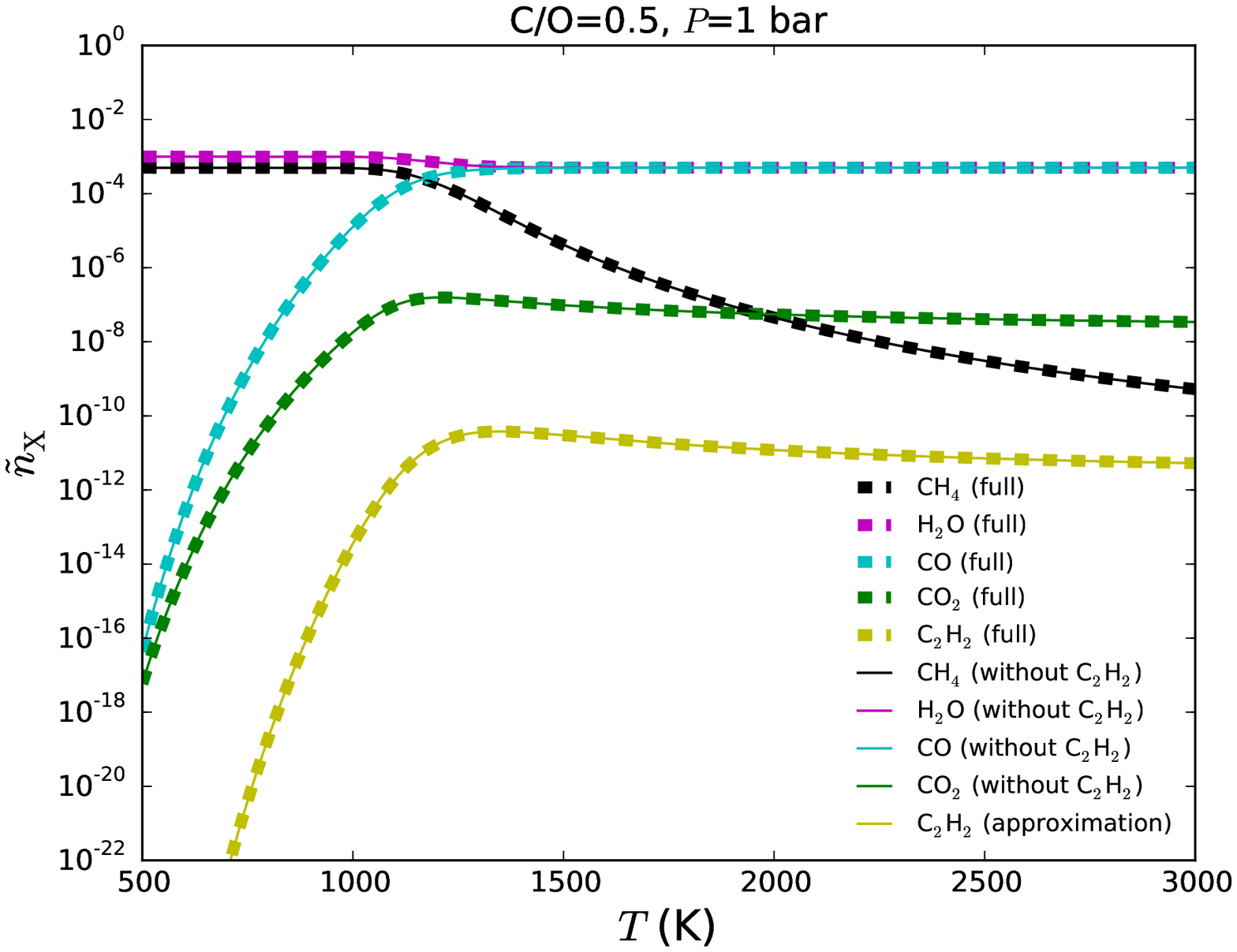}
\includegraphics[width=\columnwidth]{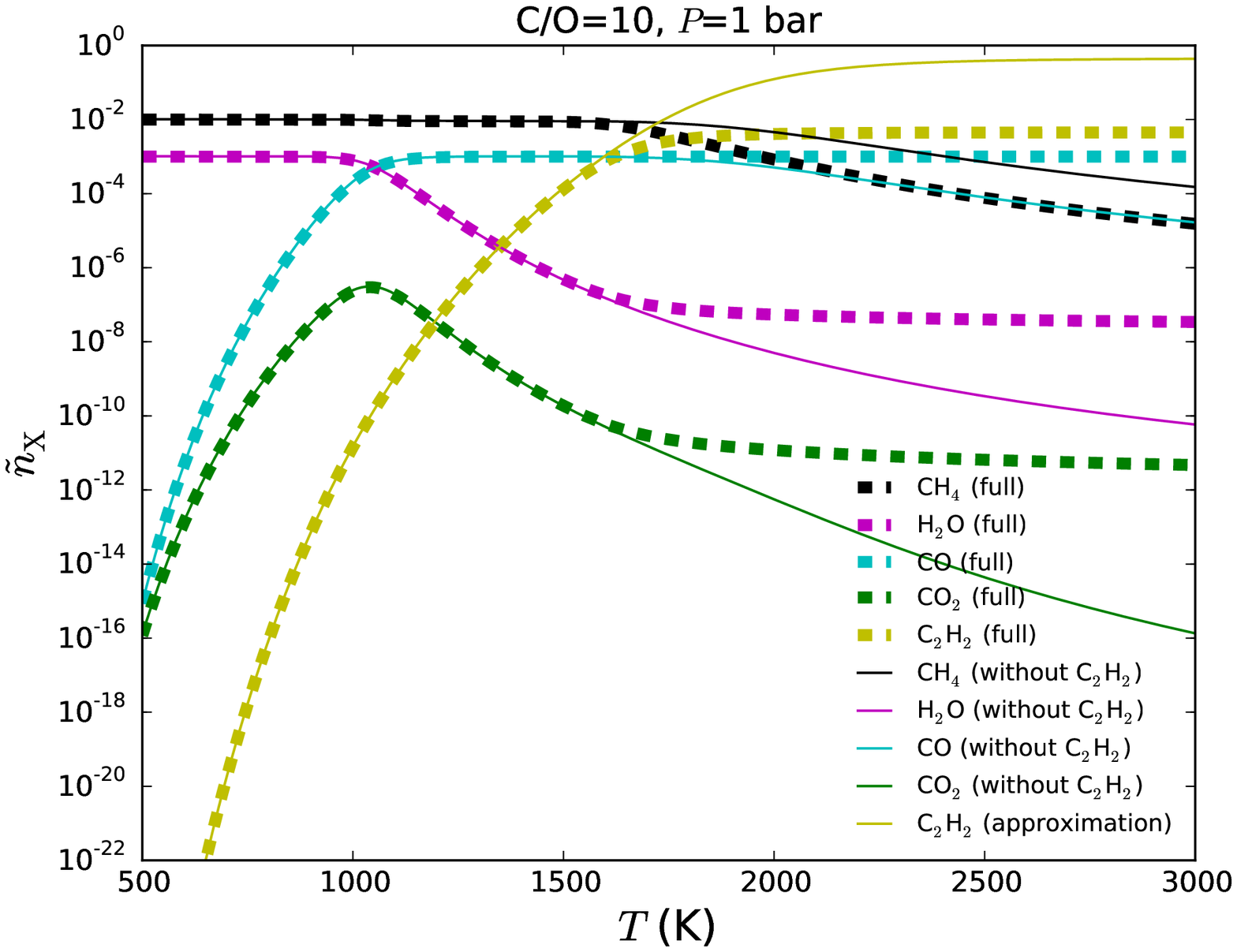}
\includegraphics[width=\columnwidth]{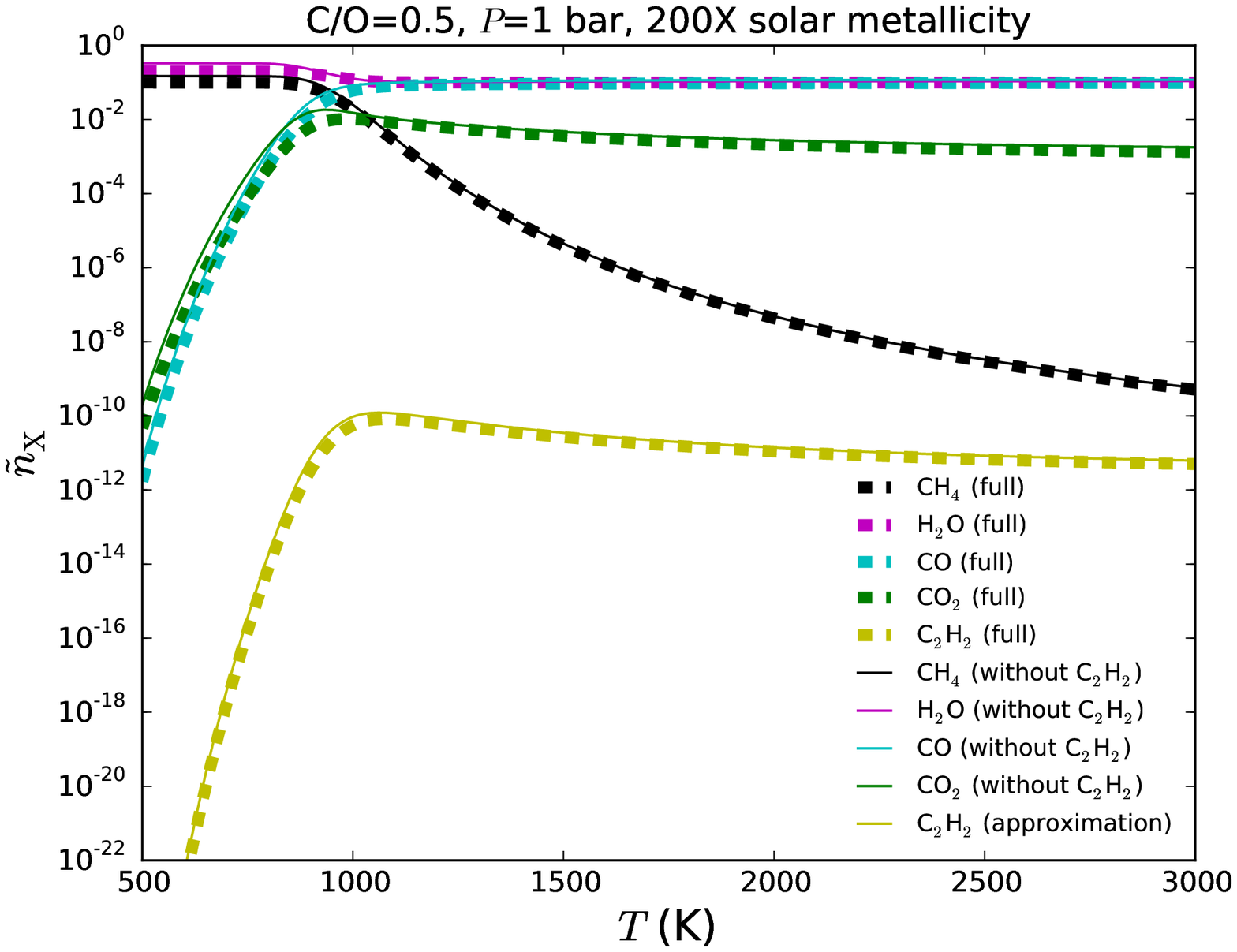}
\end{center}
%\vspace{-0.2in}
\caption{Benchmarking of our approximate carbon-rich (thick, dashed curves) versus exact carbon-poor (thin, solid curves) solutions.  The approximate solutions assume that $\tilde{n}_{\rm C}, \tilde{n}_{\rm O} \ll 1$.  When the carbon-to-oxygen ratio is low (top panel, $\mbox{C/O}=0.5$), the solutions match perfectly; the acetylene abundance is computed from $K^\prime_3 \tilde{n}_{\rm CH_4}^2$, despite it not formally being included in the carbon-poor solutions.  When it is high (middle panel, $\mbox{C/O}=10$), the abundances of methane and carbon monoxide are over- and under-estimated, respectively, because of the formal exclusion of acetylene; the estimate for acetylene, using the carbon-poor solutions, breaks down.  As the elemental abundance of oxygen becomes large ($\tilde{n}_{\rm O}=0.1$, which is a factor of 200 larger than the value at the solar photosphere), our approximate and exact solutions begin to deviate from each other (bottom panel).}
%\vspace{0.1in}
\label{fig:comparison}
\end{figure}

The main conclusion of this study is that carbon dioxide is \textit{almost always} subdominant in hot, hydrogen-dominated atmospheres, compared to carbon monoxide and water, because
\begin{equation}
\begin{split}
\frac{\tilde{n}_{\rm CO_2}}{\tilde{n}_{\rm CO}} &= \frac{\tilde{n}_{\rm H_2O}}{K^\prime_2} \sim \tilde{n}_{\rm H_2O}, \\
\frac{\tilde{n}_{\rm CO_2}}{\tilde{n}_{\rm H_2O}} &= \frac{\tilde{n}_{\rm CO}}{K^\prime_2} \sim \tilde{n}_{\rm CO}.
\end{split}
\end{equation}
We will elucidate what ``almost" means.  At this juncture, it is not so important to understand what the dimensionless function $K^\prime_2$ is---this will come later.  It is more important to note that it has no dependence on pressure and that $1/K^\prime_2 \sim 0.1$--1 for a broad range of temperatures (Figure \ref{fig:k2}).  Since we typically have $\tilde{n}_{\rm H_2O} \ll 1$ and $\tilde{n}_{\rm CO} \ll 1$ in a hydrogen- or helium-dominated atmosphere, we may conclude that, in most situations,
\begin{equation}
\frac{\tilde{n}_{\rm CO_2}}{\tilde{n}_{\rm CO}}, \frac{\tilde{n}_{\rm CO_2}}{\tilde{n}_{\rm H_2O}} \ll 1.
\label{eq:conclusion}
\end{equation}
These trends have previously been elucidated by \cite{madhu12}, \cite{koppa12}, \cite{moses13a,moses13b} and \cite{venot14}, in one form or another, using numerical calculations of chemical kinetics.  Our present study takes a complementary approach and corroborates this conclusion via a set of analytical calculations of equilibrium chemistry, spanning a broad range of temperatures, pressures, C/O values and metallicities that are representative of all of the currently observable and characterizable exoplanetary atmospheres.  (Whenever it is relevant, we point out the limitations of our analytical approach.)

Our main technical contribution is to present novel analytical solutions for chemical systems with carbon dioxide, carbon monoxide, water, methane, acetylene (C$_2$H$_2$) and molecular hydrogen, which allow us to efficiently explore vast swarths of parameter space.  We use tabulated values of the Gibbs free energy to relate the equilibrium constants to temperature and pressure.  We then fold these calculations of atmospheric chemistry with analytical models of temperature-pressure profiles to explore its effects in a wide variety of exoplanetary atmospheres.

\section{Methodology}

\subsection{Carbon-poor ($\mbox{C/O}<1$) chemistry}

We consider the conversion of methane to carbon monoxide via the net reaction \citep{bs99,lodders02,moses11},
\begin{equation}
\mbox{CH}_4 + \mbox{H}_2\mbox{O} \leftrightarrows \mbox{CO} + 3 \mbox{H}_2.
\end{equation}
This reaction alone is insufficient for modeling carbon-rich atmospheres, because hydrocarbons such as acetylene and hydrogen cyanide (HCN) are expected to appear at high temperatures \citep{madhu12,moses13a,venot15}.  We supplement it with another net reaction for producing carbon dioxide \citep{lodders02,moses11},
\begin{equation}
\mbox{CO}_2 + \mbox{H}_2 \leftrightarrows \mbox{CO} + \mbox{H}_2\mbox{O}. 
\end{equation}
If molecular hydrogen is vastly more abundant than water, then the production of carbon monoxide is favored.

The dimensional equilibrium constant of the first reaction is
\begin{equation}
K^\prime_{\rm eq} = \frac{n_{\rm CO} n_{\rm H_2}^3}{n_{\rm CH_4} n_{\rm H_2O}} = \frac{\tilde{n}_{\rm CO} n_{\rm H_2}^2}{\tilde{n}_{\rm CH_4} \tilde{n}_{\rm H_2O}},
\end{equation}
which may be normalized to obtain
\begin{equation}
K^\prime \equiv \frac{K^\prime_{\rm eq}}{n_{\rm H_2}^2}.
\end{equation}
For the second reaction, the equilibrium constant is already normalized,
\begin{equation}
K^\prime_{\rm eq,2} \equiv \frac{n_{\rm CO} n_{\rm H_2O}}{n_{\rm H_2} n_{\rm CO_2}} = \frac{\tilde{n}_{\rm CO} \tilde{n}_{\rm H_2O}}{\tilde{n}_{\rm CO_2}} \equiv K^\prime_2.
\end{equation}
Note that the number densities of the molecules marked by tildes have been normalized by $n_{\rm H_2}$, while those of the atoms will be normalized by $n_{\rm H}$.  The former are the mixing ratios, while the latter are the normalized elemental abundances.

Stoichiometric book-keeping (counting the number of atoms of each species) yields
\begin{equation}
\begin{split}
&n_{\rm CH_4} + n_{\rm CO} + n_{\rm CO_2} = n_{\rm C}, \\
&n_{\rm H_2O} + n_{\rm CO} + 2 n_{\rm CO_2} = n_{\rm O}, \\
&4n_{\rm CH_4} + 2 n_{\rm H_2O} + 2n_{\rm H_2} = n_{\rm H}.
\end{split}
\end{equation}

Manipulating these equations and using the equilibrium constants yield a cubic equation for the mixing ratio of water,
\begin{equation}
{\cal C}_3 \tilde{n}_{\rm H_2O}^3 + {\cal C}_2 \tilde{n}_{\rm H_2O}^2 + {\cal C}_1 \tilde{n}_{\rm H_2O} + {\cal C}_0 = 0,
\end{equation}
where we have defined
\begin{equation}
\begin{split}
{\cal C}_3 &= \frac{K^\prime}{K^\prime_2} \left( 2 \tilde{n}_{\rm O} - 4 \tilde{n}_{\rm C} - 1 \right) \\
{\cal C}_2 &= K^\prime \left[ 2 \left( \tilde{n}_{\rm O} - \tilde{n}_{\rm C} \right) - 1 + \frac{2}{K^\prime_2}  \left( \tilde{n}_{\rm O} - 2\tilde{n}_{\rm C} \right) \right],\\
{\cal C}_1 &= 2 \tilde{n}_{\rm O} + 4 \tilde{n}_{\rm C} - 1 + 2K^\prime \left( \tilde{n}_{\rm O} - \tilde{n}_{\rm C} \right),\\
{\cal C}_0 &= 2\tilde{n}_{\rm O}.
\end{split}
\end{equation}
Unlike when carbon dioxide is excluded, we do not obtain a quadratic equation for the mixing ratio of methane \citep{hlt15}.  We solve this cubic equation using the \texttt{polynomial.polyroots} routine in \texttt{Python}.

The rest of the mixing ratios may be obtained using
\begin{equation}
\begin{split}
\tilde{n}_{\rm CH_4} &= \frac{2 \tilde{n}_{\rm C} \left( 1 + \tilde{n}_{\rm H_2O} \right)}{1 + K^\prime \tilde{n}_{\rm H_2O} + K^\prime \tilde{n}_{\rm H_2O}^2/K^\prime_2 - 4\tilde{n}_{\rm C}}, \\
\tilde{n}_{\rm CO} &= K^\prime \tilde{n}_{\rm CH_4} \tilde{n}_{\rm H_2O}, \\
\tilde{n}_{\rm CO_2} &= \frac{\tilde{n}_{\rm CO} \tilde{n}_{\rm H_2O}}{K^\prime_2}. \\
\end{split}
\label{eq:rest}
\end{equation}
In the low-temperature limit, we obtain $\tilde{n}_{\rm H_2O} \approx 2 \tilde{n}_{\rm O}$ and $\tilde{n}_{\rm CH_4} \approx 2 \tilde{n}_{\rm C}$, such that the ratio of methane to water abundances is the carbon-to-oxygen ratio \citep{hlt15}.  At low temperatures, water and methane each sequester an extra atom of oxygen and carbon, respectively, at the expense of carbon monoxide.

\begin{table}
\label{tab:gibbs}
\begin{center}
\caption{Molar Gibbs free energies of chemical species \\ used in this study ($P_0=1$ bar)}
\begin{tabular}{lccccc}
\hline
\hline
$T$ & H$_2$O & CH$_4$ & CO & CO$_2$ & C$_2$H$_2$ \\
\hline
(K) & (kJ/mol) & (kJ/mol) & (kJ/mol) & (kJ/mol) & (kJ/mol) \\
\hline
500 & -219.051 & -32.741 & -155.414 & -394.939 & 197.452 \\
600 & -214.007 & -22.887 & -164.486 & -395.182 & 191.735 \\	
700 & -208.812 & -12.643 & -173.518 & -395.398 & 186.097 \\		
800 & -203.496 & -2.115 & -182.497 & -395.586 & 180.534	\\	
900 & -198.083 & 8.616 & -191.416 & -395.748 & 175.041 \\	
1000 & -192.590 & 19.492 & -200.275 & -395.886 & 169.607 \\		
1100 & -187.033 & 30.472 & -209.075 & -396.001 & 164.226 \\	
1200 & -181.425 & 41.524 & -217.819 & -396.098 & 158.888 \\	
1300 & -175.774 & 52.626 & -226.509	 & -396.177 & 153.588 \\		
1400 & -170.089 & 63.761 & -235.149	 & -396.240 & 148.319 \\		
1500 & -164.376 & 74.918 & -243.740 & -396.288	 & 143.078 \\		
1600 & -158.639 & 86.088 & -252.284	 & -396.323 & 137.861 \\	
1700 & -152.883 & 97.265 & -260.784	 & -396.344 & 132.665 \\		
1800 & -147.111 & 108.445 & -269.242 & -396.353 & 127.487 \\	
1900 & -141.325 & 119.624	 & -277.658 & -396.349 & 122.327 \\		
2000 & -135.528 & 130.802	 & -286.034 & -396.333 & 117.182 \\		
2100 & -129.721 & 141.975	 & -294.372 & -396.304 & 112.052 \\	
2200 & -123.905 & 153.144	 & -302.672 & -396.262 & 106.935 \\		
2300 & -118.082 & 164.308	 & -310.936 & -396.209 & 101.830 \\		
2400 & -112.252 & 175.467 & -319.165 & -396.142 & 96.738 \\	
2500 & -106.416 & 186.622	 & -327.358 & -396.062 & 91.658 \\		
2600 & -100.575 & 197.771	 & -335.517 & -395.969 & 86.589 \\		
2700 & -94.729 & 208.916 & -343.643	 & -395.862 & 81.530 \\	
2800 & -88.878 & 220.058 & -351.736	 & -395.742 & 76.483 \\		
2900 & -83.023 & 231.196 & -359.797	 & -395.609 & 71.447 \\		
3000 & -77.163 & 242.332 & -367.826	 & -395.461 & 66.421 \\	
\hline
\hline
\end{tabular}\\
%\vspace{0.05in}
Note: the molar Gibbs free energy associated with H$_2$ is 0 J mol$^{-1}$ by definition.
\end{center}
\end{table}

\begin{table}
\label{tab:gibbs2}
\begin{center}
\caption{Molar Gibbs free energies of the \\ three net reactions ($P_0=1$ bar)}
\begin{tabular}{lccc}
\hline
\hline
$T$ & $\Delta \tilde{G}_{0,1}$ & $\Delta \tilde{G}_{0,2}$ & $\Delta \tilde{G}_{0,3}$ \\
\hline
(K) & (kJ/mol) & (kJ/mol) & (kJ/mol) \\
\hline
500 & 96.378 & 20.474 & 262.934 \\
600 & 72.408 & 16.689 & 237.509 \\		
700 & 47.937 & 13.068 & 211.383 \\		
800 & 23.114 & 9.593 & 184.764 \\		
900 & -1.949 & 6.249 & 157.809 \\		
1000 & -27.177 & 3.021 & 130.623 \\		
1100 & -52.514 & -0.107 & 103.282 \\		
1200 & -77.918 & -3.146 & 75.840 \\		
1300 & -103.361 & -6.106 & 48.336 \\		
1400 & -128.821 & -8.998 & 20.797 \\		
1500 & -154.282 & -11.828 & -6.758 \\		
1600 & -179.733 & -14.600 & -34.315 \\		
1700 & -205.166 & -17.323 & -61.865 \\		
1800 & -230.576 & -20.000 & -89.403 \\		
1900 & -255.957 & -22.634 & -116.921 \\		
2000 & -281.308 & -25.229 & -144.422 \\	
2100 & -306.626 & -27.789 & -171.898 \\	
2200 & -331.911 & -30.315 & -199.353 \\		
2300 & -357.162 & -32.809 & -226.786 \\		
2400	 & -382.38	 & -35.275	 & -254.196 \\		
2500	 & -407.564 & -37.712 & -281.586 \\			
2600	 & -432.713 & -40.123 & -308.953 \\		
2700	 & -457.830 & -42.509 & -336.302 \\		
2800	 & -482.916 & -44.872 & -363.633 \\			
2900	 & -507.97 & -47.211 & -390.945 \\		
3000	 & -532.995 & -49.528 & -418.243 \\			
\hline
\hline
\end{tabular}\\
%\vspace{0.05in}
\end{center}
\end{table}

\subsection{Carbon-rich ($\mbox{C/O}>1$) chemistry}

Following \cite{hlt15}, we include acetylene as a proxy for all of the hydrocarbons that may form at high temperatures and in carbon-rich situations, via the net reaction \citep{lodders02,moses11},
\begin{equation}
2\mbox{CH}_4 \leftrightarrows \mbox{C}_2\mbox{H}_2 + 3 \mbox{H}_2.
\end{equation}
Its equilibrium constant is
\begin{equation}
K^\prime_{\rm eq,3} = \frac{n_{\rm C_2H_2} n_{\rm H_2}^3}{n_{\rm CH_4}^2} = \frac{\tilde{n}_{\rm C_2H_2} n_{\rm H_2}^2}{\tilde{n}_{\rm CH_4}^2},
\end{equation}
which we also normalize,
\begin{equation}
K^\prime_3 \equiv \frac{K^\prime_{\rm eq,3}}{n_{\rm H_2}^2}.
\end{equation}

Stoichiometric book-keeping is generalized to
\begin{equation}
\begin{split}
&n_{\rm CH_4} + n_{\rm CO} + n_{\rm CO_2} + 2 n_{\rm C_2H_2} = n_{\rm C}, \\
&n_{\rm H_2O} + n_{\rm CO} + 2 n_{\rm CO_2} = n_{\rm O}, \\
&4n_{\rm CH_4} + 2 n_{\rm H_2O} + 2n_{\rm H_2} + 2 n_{\rm C_2H_2} = n_{\rm H}.
\end{split}
\end{equation}

\begin{figure*}%[!h]
\begin{center}
%\vspace{-0.2in}
\includegraphics[width=\columnwidth]{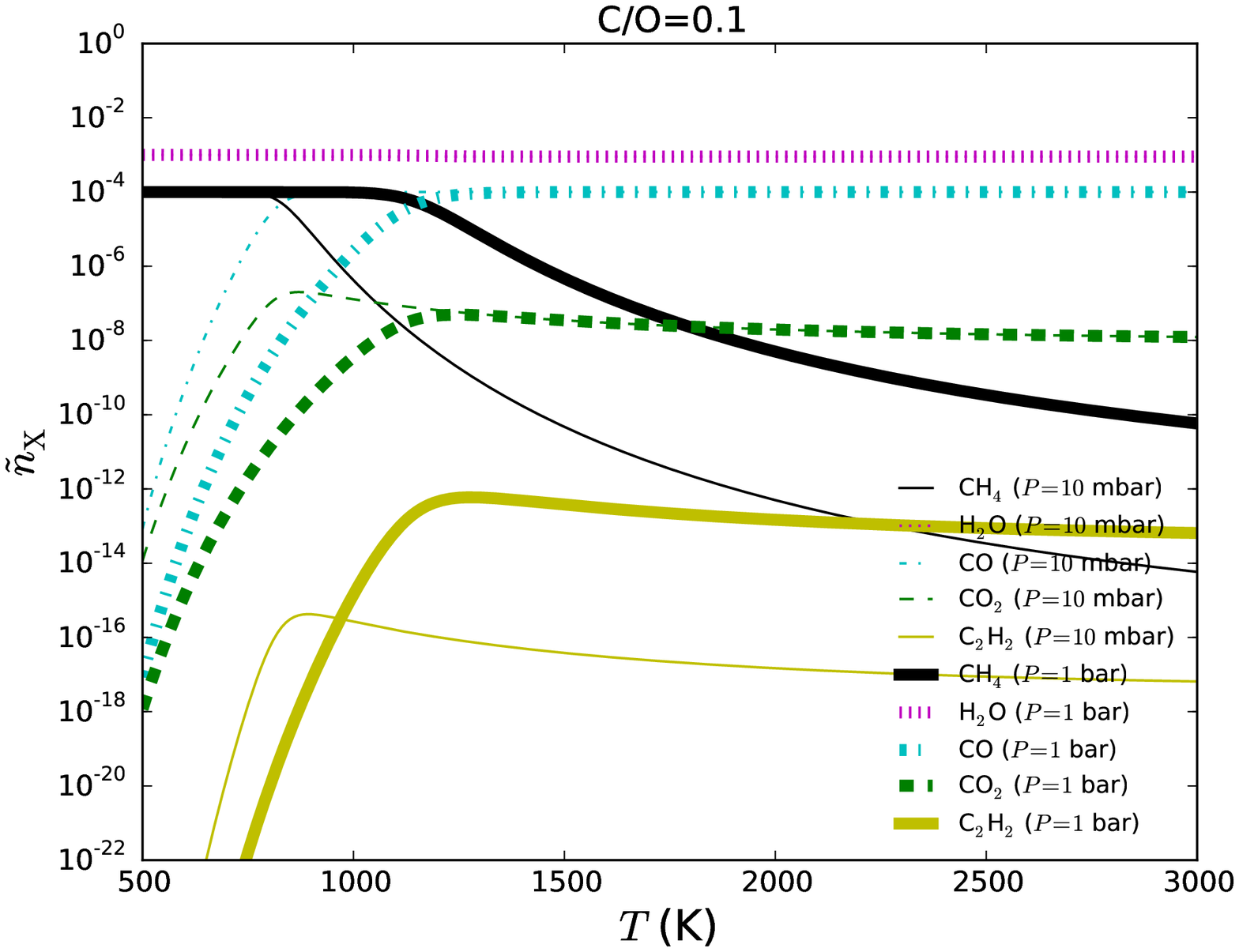}
\includegraphics[width=\columnwidth]{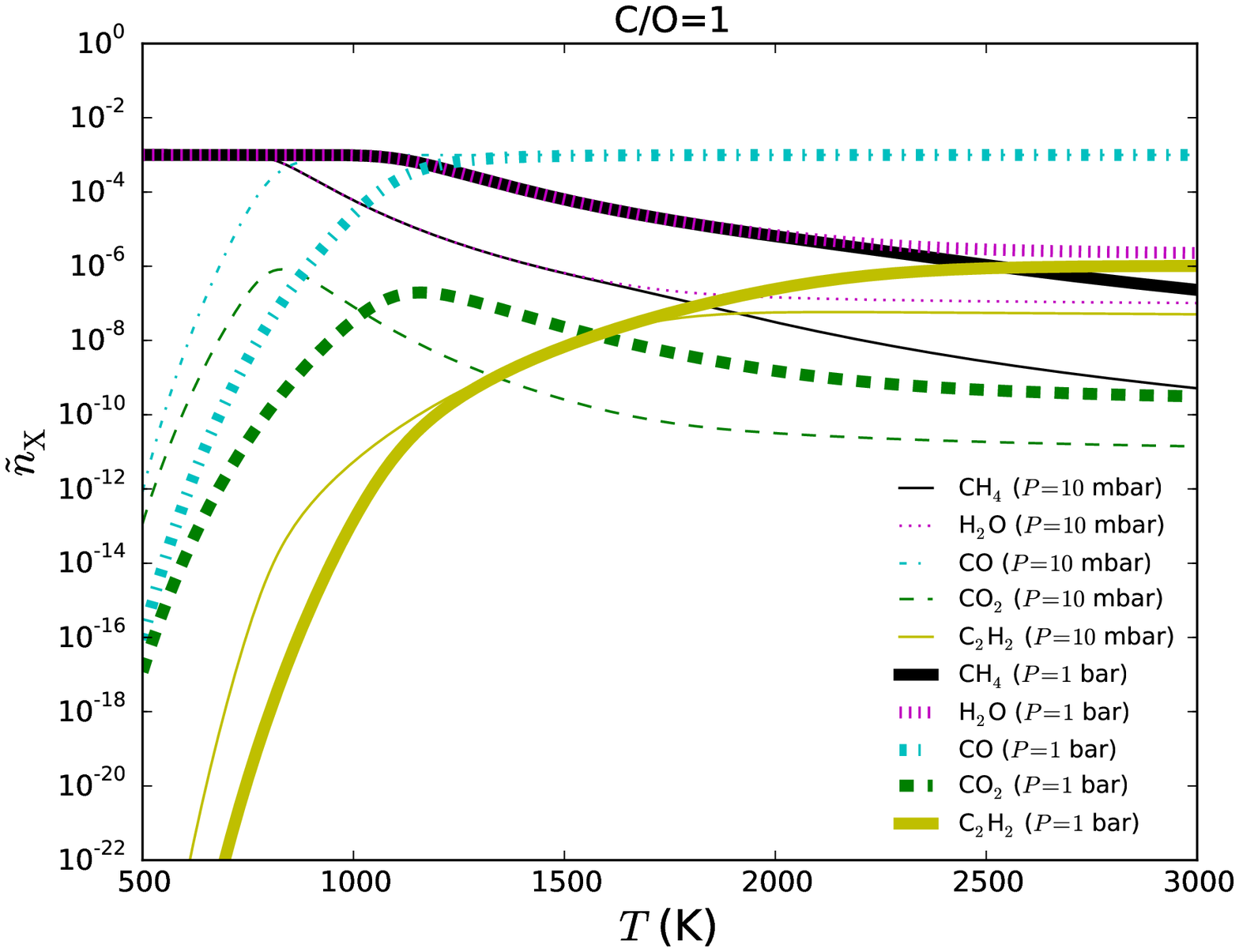}
\includegraphics[width=\columnwidth]{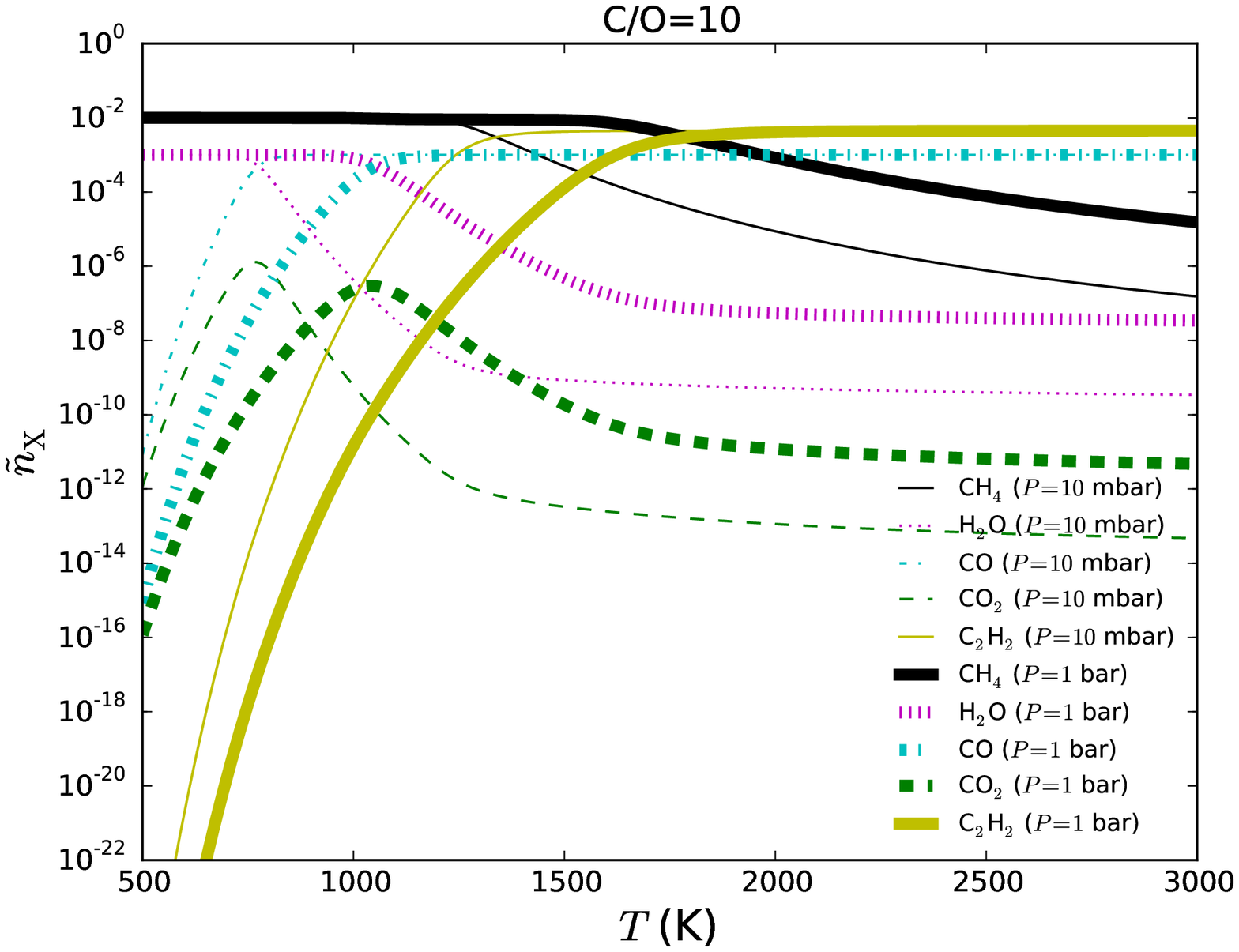}
\includegraphics[width=\columnwidth]{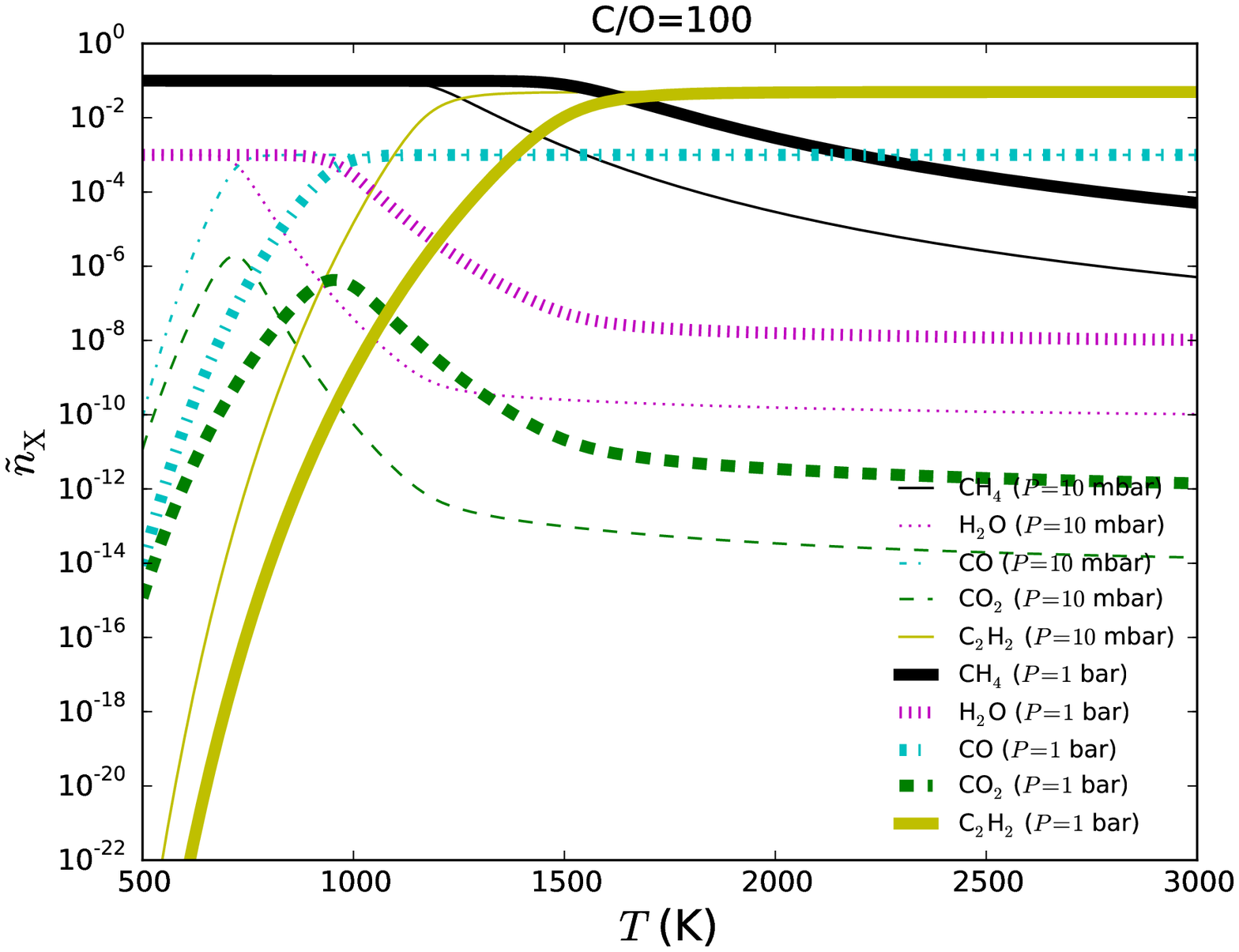}
\end{center}
%\vspace{-0.2in}
\caption{Mixing ratios of molecules as a function of temperature.  We have assumed $\tilde{n}_{\rm O}=5 \times 10^{-4}$.  Top-left panel: sub-solar carbon abundance ($\mbox{C/O}=0.1$).  Top-right panel: $\mbox{C/O}=1$.  Bottom-left panel: $\mbox{C/O}=10$.  Bottom-right panel: $\mbox{C/O}=100$.}
%\vspace{0.1in}
\label{fig:abundances}
\end{figure*}

Instead of a cubic equation for the mixing ratio of water, we obtain a pair of \textit{coupled} quadratic equations for the mixing ratios of water and methane,
\begin{equation}
\begin{split}
&\frac{K^\prime \tilde{n}_{\rm H_2O}^2 \tilde{n}_{\rm CH_4}}{K^\prime_2} + 2 K^\prime_3 \tilde{n}_{\rm CH_4}^2 \left( 1 - \tilde{n}_{\rm C} \right) - 2 \tilde{n}_{\rm C} \tilde{n}_{\rm H_2O} \\
&+ \tilde{n}_{\rm CH_4} + K^\prime \tilde{n}_{\rm H_2O} \tilde{n}_{\rm CH_4} - 4\tilde{n}_{\rm C} \tilde{n}_{\rm CH_4} - 2\tilde{n}_{\rm C} = 0, \\
&\frac{2K^\prime \tilde{n}_{\rm H_2O}^2 \tilde{n}_{\rm CH_4}}{K^\prime_2} - 2 K^\prime_3 \tilde{n}_{\rm O} \tilde{n}_{\rm CH_4}^2 + \tilde{n}_{\rm H_2O} \left( 1 - 2 \tilde{n}_{\rm O} \right)  \\
&+ K^\prime \tilde{n}_{\rm H_2O} \tilde{n}_{\rm CH_4} - 4\tilde{n}_{\rm O} \tilde{n}_{\rm CH_4} - 2\tilde{n}_{\rm O} = 0.
\end{split}
\label{eq:coupled}
\end{equation}
The preceding pair of equations looks deceptively simple, but it is actually difficult to solve numerically as the mixing ratios span more than 20 orders of magnitude in value across the range of temperatures and pressures we are interested in.  It is challenging to maintain numerical stability across such an enormous range of values.  Taking an analytical approximation is tricky, because it is difficult to judge which terms to drop or neglect.  It turns out that we may employ an algebraic trick if we recast the equations in (\ref{eq:coupled}) in terms of $\tilde{n}_{\rm CO}$, rather than $\tilde{n}_{\rm H_2O}$, which allows us to combine them into a single equation that is essentially a quadratic equation for the mixing ratio of CO,
\begin{equation}
\begin{split}
&\frac{\tilde{n}_{\rm CO}^2}{K^\prime K^\prime_2} + \frac{\tilde{n}_{\rm CO}}{K^\prime} \left( 1 + 2 \tilde{n}_{\rm C} - 2 \tilde{n}_{\rm O} \right) - 2K^\prime_3 \tilde{n}_{\rm CH_4}^3 \left( 1 - \tilde{n}_{\rm C} + \tilde{n}_{\rm O} \right) \\
&-  \tilde{n}_{\rm CH_4}^2 - 2 \tilde{n}_{\rm CH_4} \left( \tilde{n}_{\rm O} - \tilde{n}_{\rm C}  \right) \left( 2\tilde{n}_{\rm CH_4} + 1 \right) = 0.
\end{split}
\end{equation}
Notice how the mixing ratios of carbon monoxide and methane are no longer coupled to each other within the same equation---there are no ``mixed" terms, unlike for each equation in (\ref{eq:coupled}) between water and methane.  This property has the virtue that we may cleanly take the approximation $\tilde{n}_{\rm C}, \tilde{n}_{\rm O} \ll 1$ and end up with a relatively simple expression for the mixing ratio of carbon monoxide,
\begin{equation}
\tilde{n}_{\rm CO} \approx K^\prime \tilde{n}_{\rm CH_4} \left[ 2 K^\prime_3 \tilde{n}_{\rm CH_4}^2 + \tilde{n}_{\rm CH_4} + 2 \left( \tilde{n}_{\rm O} - \tilde{n}_{\rm C} \right) \right].
\end{equation}
Despite not being exact, the preceding expression retains generality, because the only assumption we have made so far is that the elemental abundances are small compared to hydrogen.  It is less obvious how to take the $\tilde{n}_{\rm C}, \tilde{n}_{\rm O} \ll 1$ approximation directly using the equations in (\ref{eq:coupled}).

The expression for $\tilde{n}_{\rm CO}$ leads to an approximate expression for the mixing ratio of water,
\begin{equation}
\tilde{n}_{\rm H_2O} \approx 2 K^\prime_3 \tilde{n}_{\rm CH_4}^2 + \tilde{n}_{\rm CH_4} + 2 \left( \tilde{n}_{\rm O} - \tilde{n}_{\rm C} \right),
\end{equation}
and a quintic equation for the mixing ratio of methane,
\begin{equation}
\sum_{i=0}^5 {\cal A}_i \tilde{n}_{\rm CH_4}^i \approx 0,
\end{equation}
where the coefficients are
\begin{equation}
\begin{split}
{\cal A}_5 &= \frac{8 K^\prime K^{\prime 2}_3}{K^\prime_2}, \\
{\cal A}_4 &= \frac{8 K^\prime K^\prime_3}{K^\prime_2}, \\
{\cal A}_3 &= \frac{2 K^\prime}{K^\prime_2} \left[ 1 + 8 K^\prime_3 \left( \tilde{n}_{\rm O} - \tilde{n}_{\rm C} \right) \right] + 2 K^\prime K^\prime_3, \\
{\cal A}_2 &= \frac{8 K^\prime}{K^\prime_2}\left( \tilde{n}_{\rm O} - \tilde{n}_{\rm C} \right) + 2 K^\prime_3 + K^\prime, \\
{\cal A}_1 &= \frac{8 K^\prime}{K^\prime_2}\left( \tilde{n}_{\rm O} - \tilde{n}_{\rm C} \right)^2 + 1 + 2 K^\prime \left( \tilde{n}_{\rm O} - \tilde{n}_{\rm C} \right), \\
{\cal A}_0 &= - 2 \tilde{n}_{\rm C}.
\end{split}
\end{equation}
The mixing ratios of CO and CO$_2$ may be obtained using the second and third equations in (\ref{eq:rest}), respectively, while we have $\tilde{n}_{\rm C_2H_2} = K^\prime_3 \tilde{n}_{\rm CH_4}^2$.  An indication that this equation is correct comes from the fact that it automatically yields $\tilde{n}_{\rm CH_4} \approx 2 \tilde{n}_{\rm C}$ when all of the normalized equilibrium constants vanish (i.e., the low-temperature limit).  Figure \ref{fig:comparison} shows the benchmarking of our carbon-rich versus carbon-poor solutions, which match perfectly when $\mbox{C/O}<1$.

This (approximate) quintic equation that governs the mixing ratio of methane is a novel result that generalizes the work of \cite{bs99} and \cite{hlt15}.  \cite{bs99} excluded both acetylene and carbon dioxide in the analytical solution listed in their appendix, while \cite{hlt15} did not consider carbon dioxide.

\subsection{Adopting the chemist's convention for thermodynamic quantities}

Let the ${\cal R}_{\rm univ}$ be the universal gas constant, $\mu$ be the mean molecular weight, $m = \mu m_{\rm u}$ be the mean molecular mass, $m_{\rm u}$ is the atomic mass unit and $k_{\rm B}={\cal R}_{\rm univ}/N_{\rm A}$ be the Boltzmann constant.  If we denote the mass density by $\rho$, the volume by $V$, the temperature by $T$, the number of particles by $N$ and the number density by $n$, then the ideal gas law may be expressed as
\begin{equation}
P = n k_{\rm B} T = \frac{N {\cal R}_{\rm univ} T}{V N_{\rm A}},
\end{equation}
where $N_{\rm A} = 6.02214129 \times 10^{23}$ mol$^{-1}$ is Avogrado's constant.  A mole of matter contains exactly $N_{\rm A}$ particles.  By definition, $m_{\rm u} N_{\rm A} \equiv 1$ g mol$^{-1}$.  We write the ideal gas law in this way, because we wish to express the universal gas constant in mks units as ${\cal R}_{\rm univ} = 8.3144621$ J K$^{-1}$ mol$^{-1}$ in order to utilize the data from the thermodynamic databases.  There is some confusion over the definition of the universal gas constant, because one may only switch between its erg K$^{-1}$ mol$^{-1}$ and erg K$^{-1}$ g$^{-1}$ forms, without paying a unit conversion penalty, when dealing with cgs units.

\begin{figure*}%[!h]
\begin{center}
%\vspace{-0.2in}
\includegraphics[width=\columnwidth]{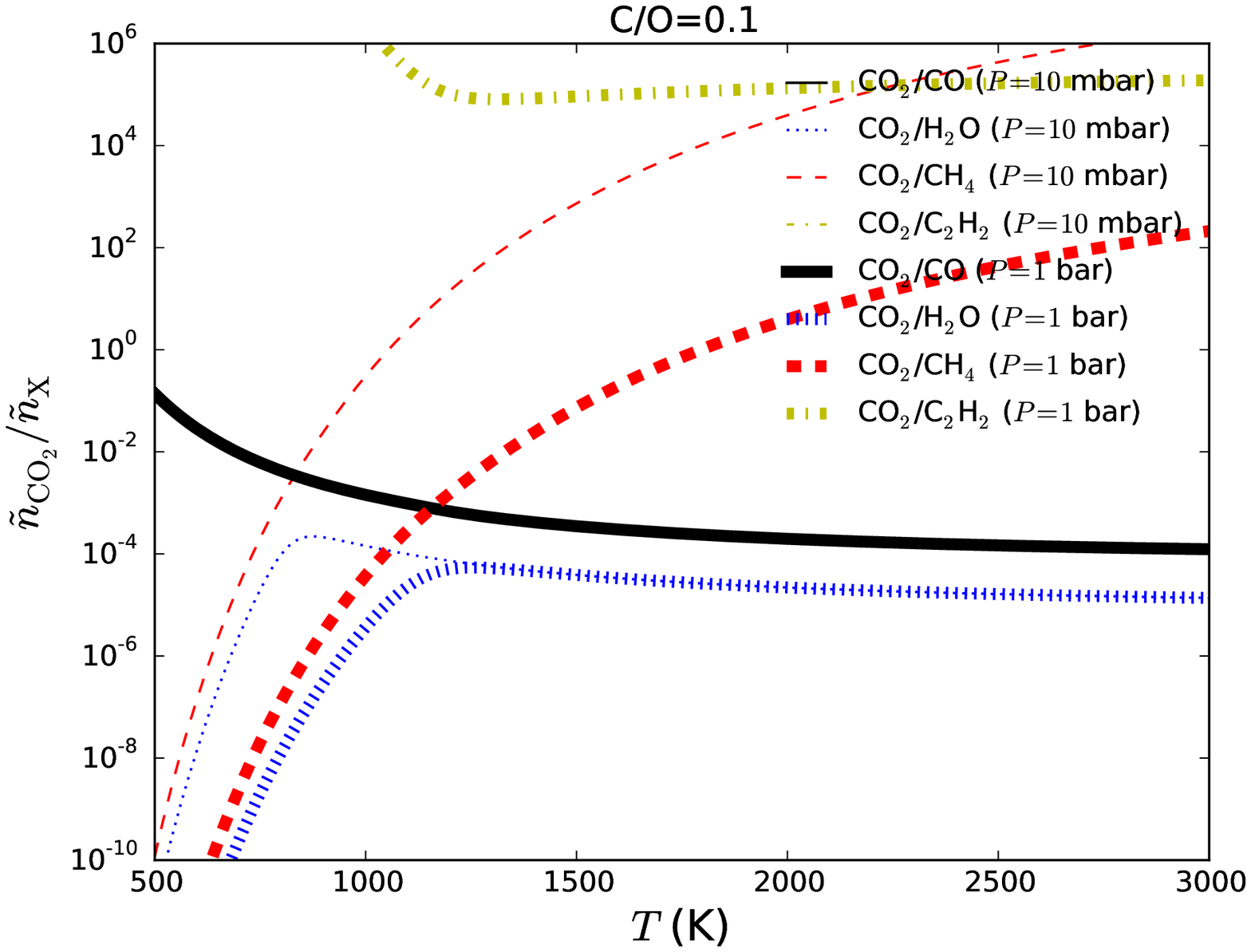}
\includegraphics[width=\columnwidth]{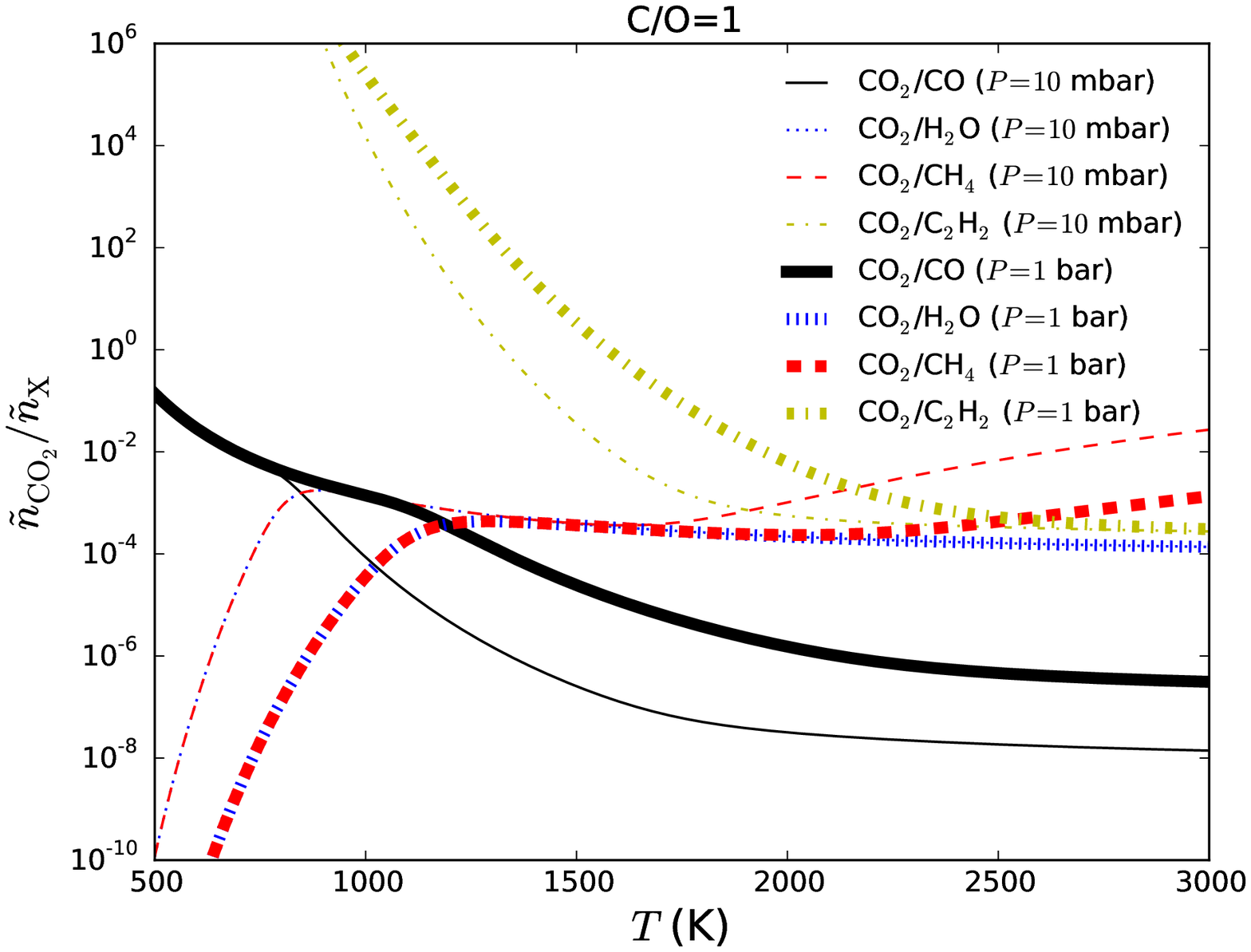}
\includegraphics[width=\columnwidth]{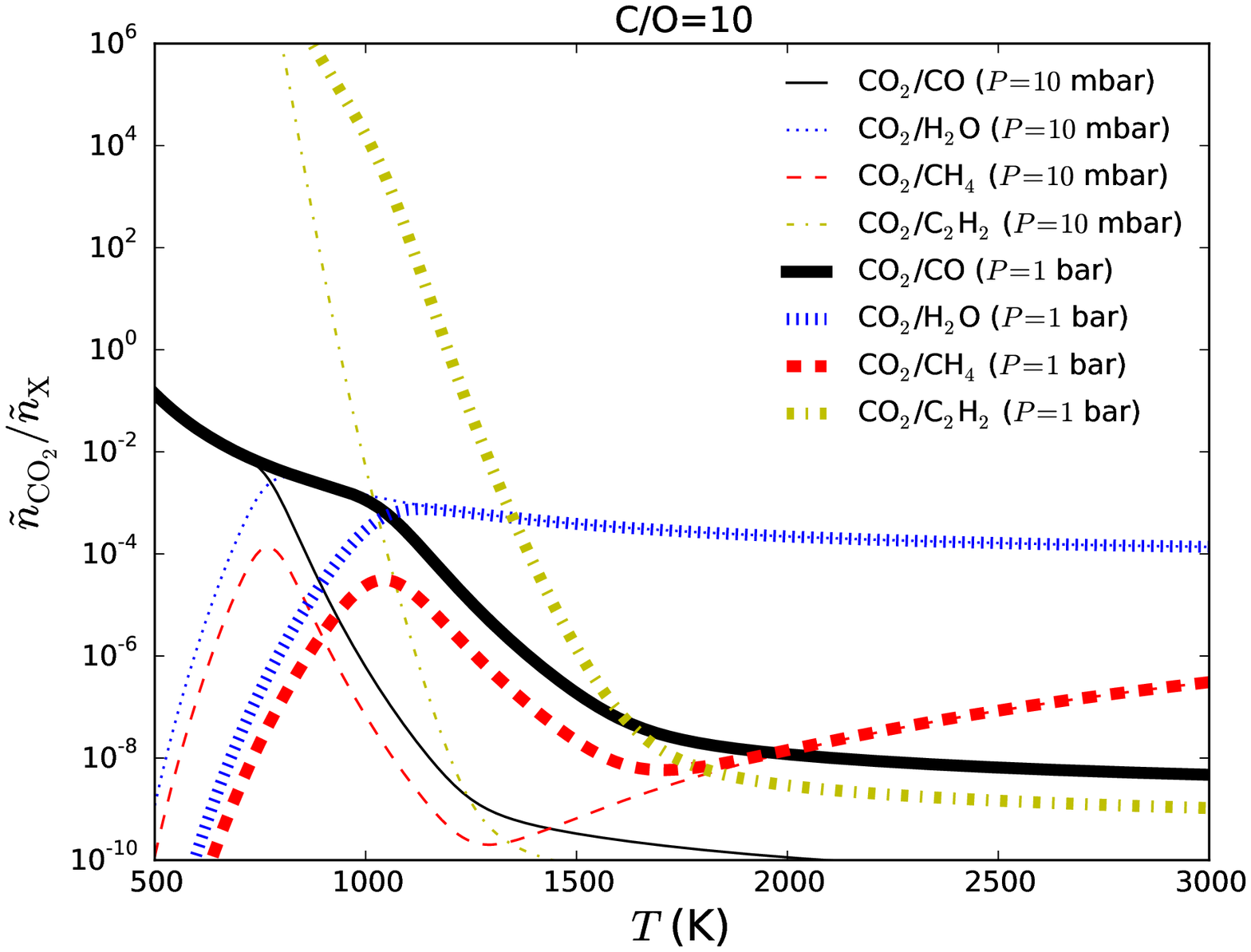}
\includegraphics[width=\columnwidth]{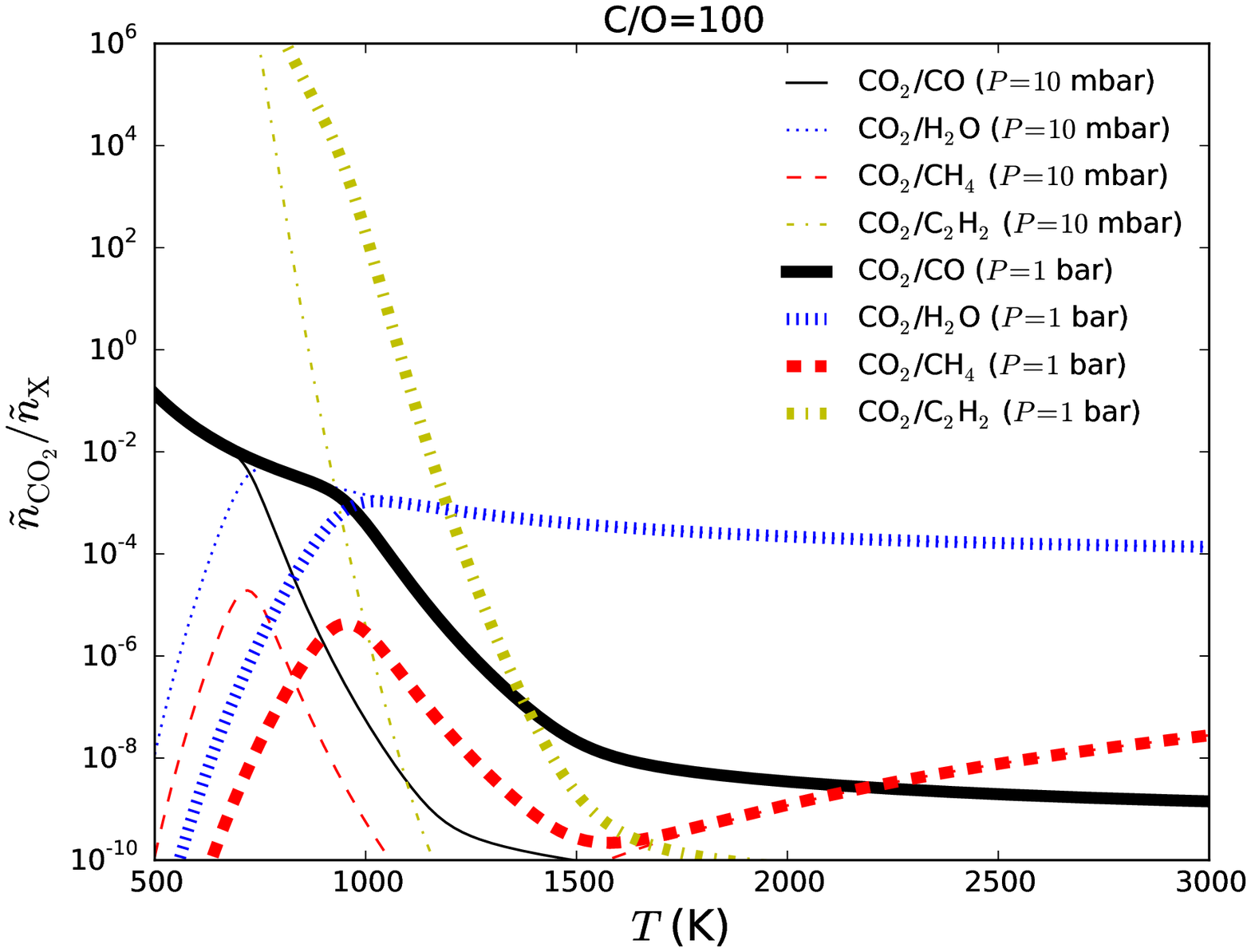}
\end{center}
%\vspace{-0.2in}
\caption{Ratio of the mixing ratios of carbon dioxide to the other molecules.  We have assumed $\tilde{n}_{\rm O}=5 \times 10^{-4}$.  Top-left panel: sub-solar carbon abundance ($\mbox{C/O}=0.1$).  Top-right panel: $\mbox{C/O}=1$.  Bottom-left panel: $\mbox{C/O}=10$.  Bottom-right panel: $\mbox{C/O}=100$.}
%\vspace{0.1in}
\label{fig:ratios}
\end{figure*}

In a departure from the formalism of \cite{hlt15}, we write $G$ as the Gibbs free energy, rather than the specific Gibbs free energy.  It follows that the scaling relation involving it becomes
\begin{equation}
\frac{G}{N/N_{\rm A}} = \frac{G_0}{N/N_{\rm A}} + {\cal R}_{\rm univ} T ~\ln{\left(\frac{P}{P_0}\right)},
\end{equation}
where $G_0$ is the Gibbs free energy at the reference pressure ($P_0$).  Typically, it is the molar Gibbs free energy at a reference pressure ($\tilde{G}_0 \equiv G_0 N_{\rm A}/N$), and not $G_0$, that is tabulated in thermodynamic databases.  It has units of J mol$^{-1}$.

\subsection{Equilibrium constants, Gibbs free energy, temperature and pressure}
\label{subsect:gibbs}

We need to distinguish between the different definitions of the equilibrium constant \citep{vm11,hlt15}.  The \textit{dimensionless} equilibrium constant is (e.g., \citealt{jacobson})
\begin{equation}
K_{\rm eq} = \exp{\left(-\frac{\Delta \tilde{G}_{0,1}}{{\cal R}_{\rm univ} T} \right)},
\end{equation}
where $\Delta \tilde{G}_{0,1} \equiv \Delta G_{0,1} N_{\rm A} / N$ and $\Delta G_{0,1}$ is the change in Gibbs free energy going from the reactants to the products, at the reference pressure, associated with the first net reaction.  $K_{\rm eq}$ should not be confused with the \textit{dimensional} equilibrium constants we have used so far, which have been rendered dimensionless by dividing by $n^2_{\rm H_2}$.  

If we focus on the first net reaction, then the dimensionless ($K_{\rm eq}$), dimensional ($K^\prime_{\rm eq}$) and normalized ($K^\prime$) equilibrium constants are related by \citep{hlt15}
\begin{equation}
K_{\rm eq}^\prime = n_0^2 K_{\rm eq} = n^2_{\rm H_2} K^\prime,
\label{eq:two_keq}
\end{equation}
where $n_0 = P_0/k_{\rm B} T$ is the number density corresponding to the reference pressure and $P = n_{\rm H_2} k_{\rm B} T$.  Thus, we obtain 
\begin{equation}
K^\prime = \left( \frac{P_0}{P} \right)^2 \exp{\left(-\frac{\Delta \tilde{G}_{0,1}}{{\cal R}_{\rm univ} T} \right)}.
\end{equation}
For the reaction involving carbon dioxide, we have
\begin{equation}
K^\prime_2 = \exp{\left(-\frac{\Delta \tilde{G}_{0,2}}{{\cal R}_{\rm univ} T} \right)}.
\end{equation}
For the reaction involving acetylene, we have
\begin{equation}
K^\prime_3 = \left( \frac{P_0}{P} \right)^2 \exp{\left(-\frac{\Delta \tilde{G}_{0,3}}{{\cal R}_{\rm univ} T} \right)}.
\end{equation}
Here, the pressure $P$ is interpreted as the total pressure of the atmosphere, which is exerted mostly by H$_2$ in hydrogen-dominated atmospheres.

Table 1 lists calculations or measurements of the molar Gibbs free energies of the molecules involved in the present study, taken from the NIST-JANAF database\footnote{\texttt{http://kinetics.nist.gov/janaf/}}.  Table 2 lists the change in molar Gibbs free energy, at the reference pressure of $P_0=1$ bar, for all three net reactions, which allow us to relate $K^\prime$, $K^\prime_2$ and $K^\prime_3$ to $T$ and $P$ (Figure \ref{fig:keq}).

\subsection{Temperature-pressure profiles of model atmospheres}

\begin{figure}%[!h]
\begin{center}
%\vspace{-0.2in}
\includegraphics[width=\columnwidth]{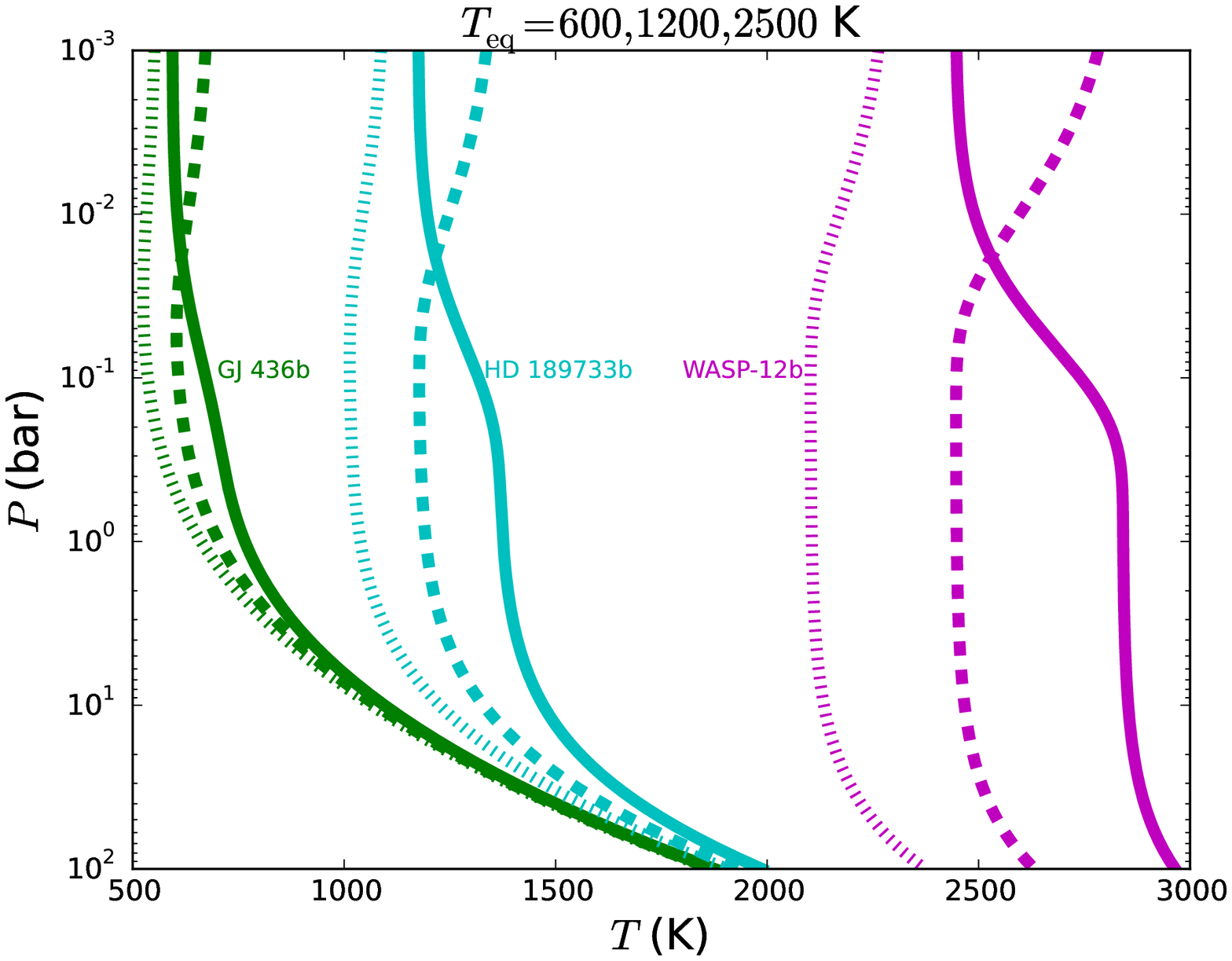}
\includegraphics[width=\columnwidth]{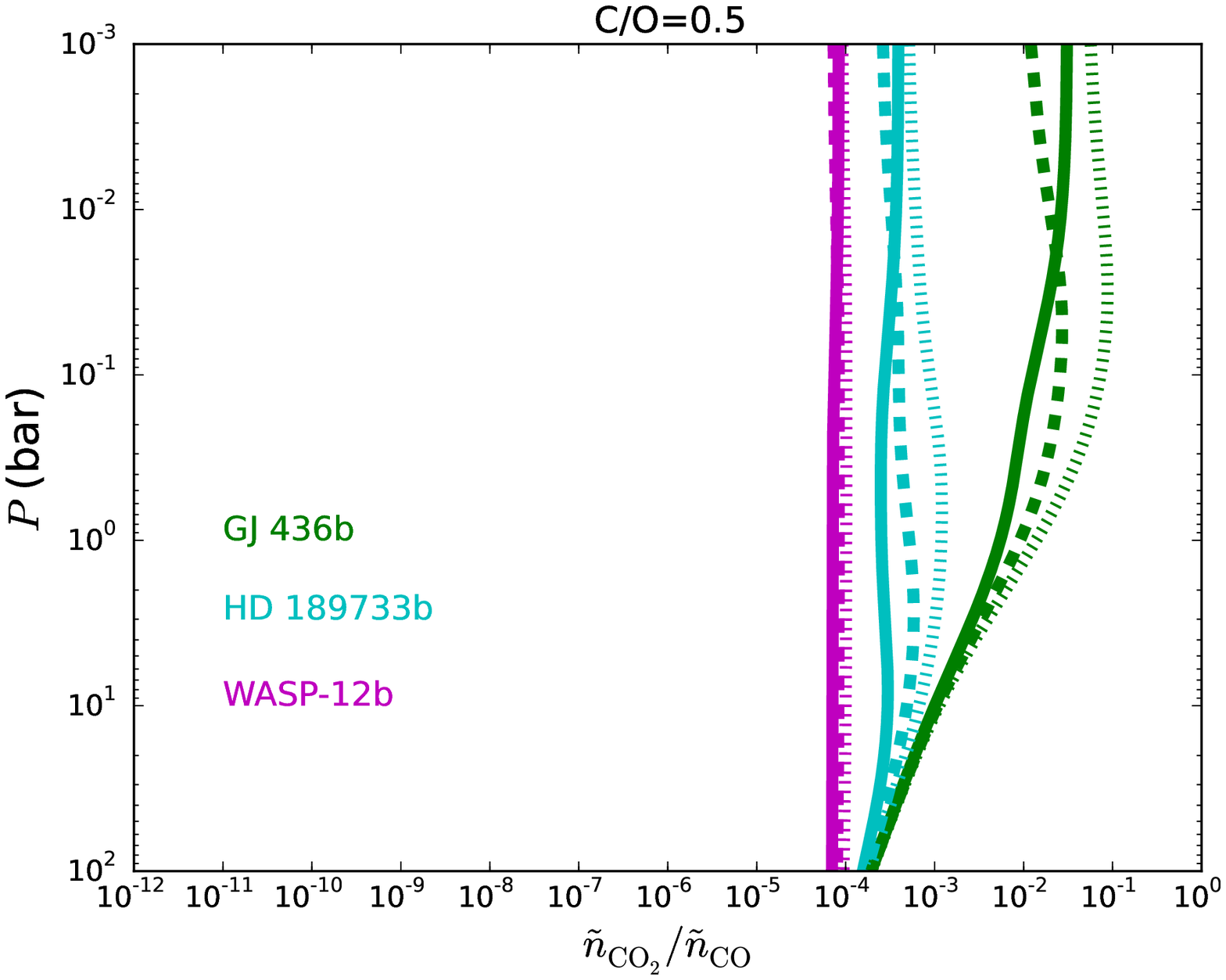}
\includegraphics[width=\columnwidth]{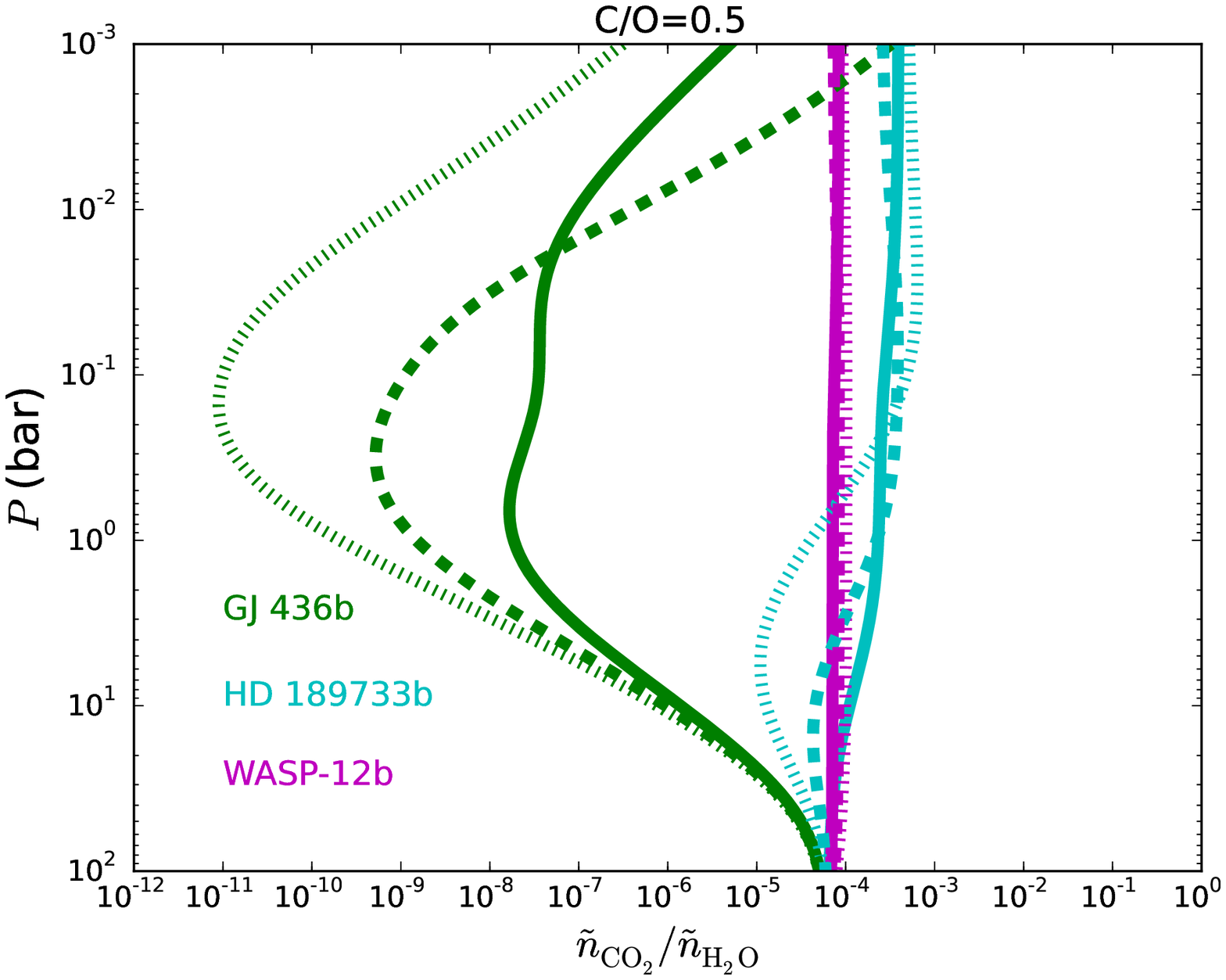}
\end{center}
%\vspace{-0.2in}
\caption{Temperature-pressure profiles of model atmospheres with equilibrium temperatures of $T_{\rm eq} = 600$, 1200 and 2500 K, representative of GJ 436b, HD 189733b and WASP-12b, respectively (top panel).  For each equilibrium temperature, we vary the Bond albedo and visible opacity of the atmosphere.  We then calculate the abundance of carbon dioxide relative to carbon monoxide (middle panel) and water (bottom panel).  In all panels, the solid curve is the fiducial model for each $T_{\rm eq}$ (zero albedo, no temperature inversion).  The dotted curve is for $A_{\rm B}=0.5$, while the dashed curve introduces a temperature inversion by increasing the visible opacity by a factor of 4.  We assume a solar abundance of elements ($\tilde{n}_{\rm O}=5 \times 10^{-4}$ and $\mbox{C/O}=0.5$).}
%\vspace{0.1in}
\label{fig:tp}
\end{figure}

To apply our calculations of equilibrium chemistry to atmospheres, we need to know their thermal structures.  To this end, we employ the analytical temperature-pressure profiles of \cite{hhps12} and \cite{hml14}, which generalized the pure-absorption models of \cite{guillot10} to include non-isotropic scattering in both the visible and infrared range of wavelengths.  These models require a small number of input parameters: the equilibrium temperature ($T_{\rm eq}$), the interior/internal temperature ($T_{\rm int}$), the visible opacity ($\kappa_{\rm vis}$), the infrared opacity ($\kappa_{\rm IR}$) and the Bond albedo ($A_{\rm B}$).  Two of these parameters are directly observable or inferable quantities ($T_{\rm eq}$ and $A_{\rm B}$).  In principle, the opacities may be the outcome of a retrieval calculation.  The internal temperature is a quantity that is unconstrained by the observations for transiting exoplanets, but we will adopt a finite value to include its effect.  We do not explore the effects of collision-induced absorption, scattering by large particles in the infrared or Gaussian cloud decks, as these are secondary effects.

It is important to note that these temperature-pressure profiles are formal solutions of the radiative transfer equation and thus enforce local \textit{and} global energy conservation---with the former being radiative equilibrium---\textit{by construction} (see Appendix).  They are not ad hoc fitting or parametric functions, which are commonly used in retrieval models.  Rather, they are first-principle calculations with simplifying approximations taken, the most major of which is that starlight and thermal emission are grouped into separate wavebands.

\section{Results}

\begin{figure*}%[!h]
\begin{center}
%\vspace{-0.2in}
\includegraphics[width=\columnwidth]{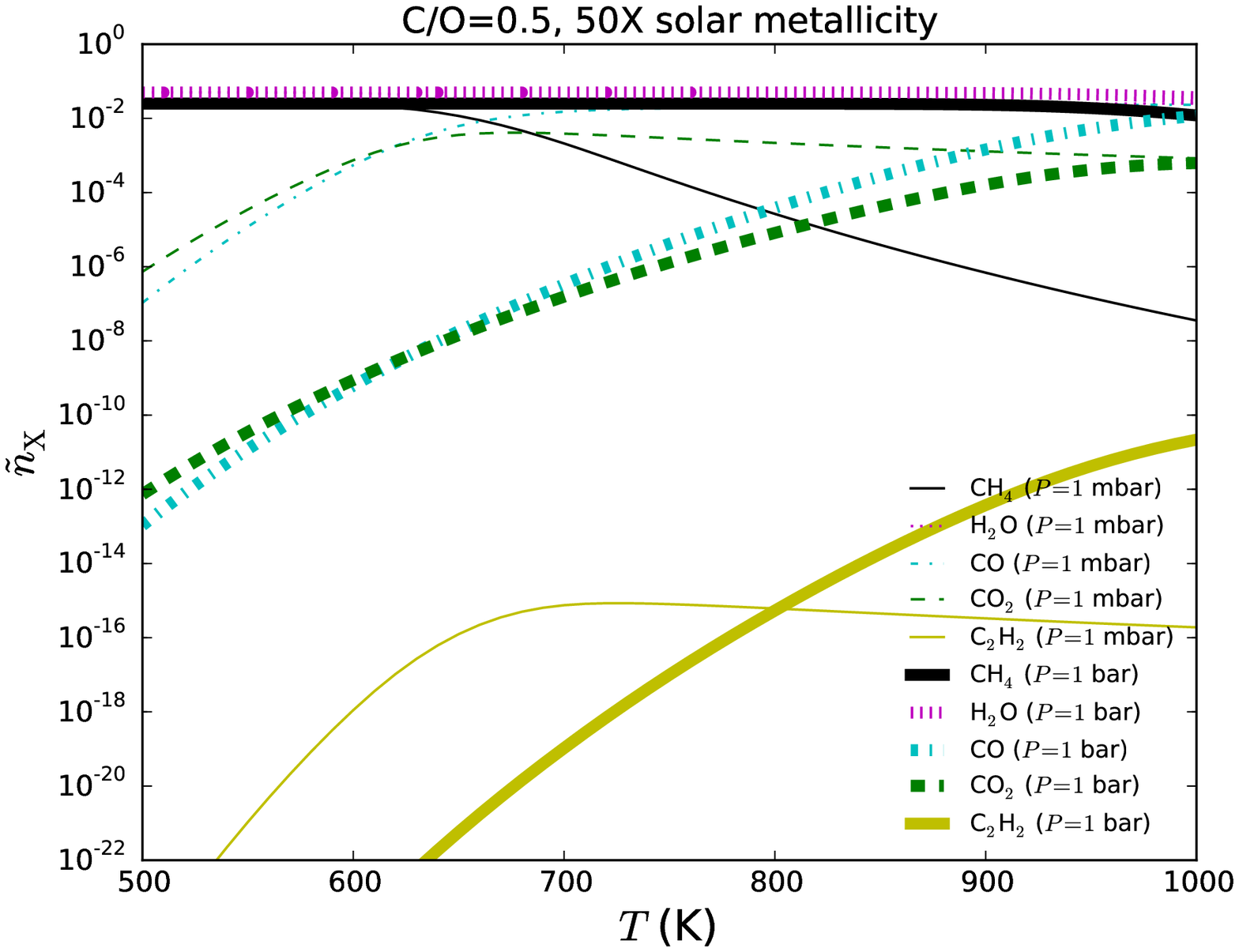}
\includegraphics[width=\columnwidth]{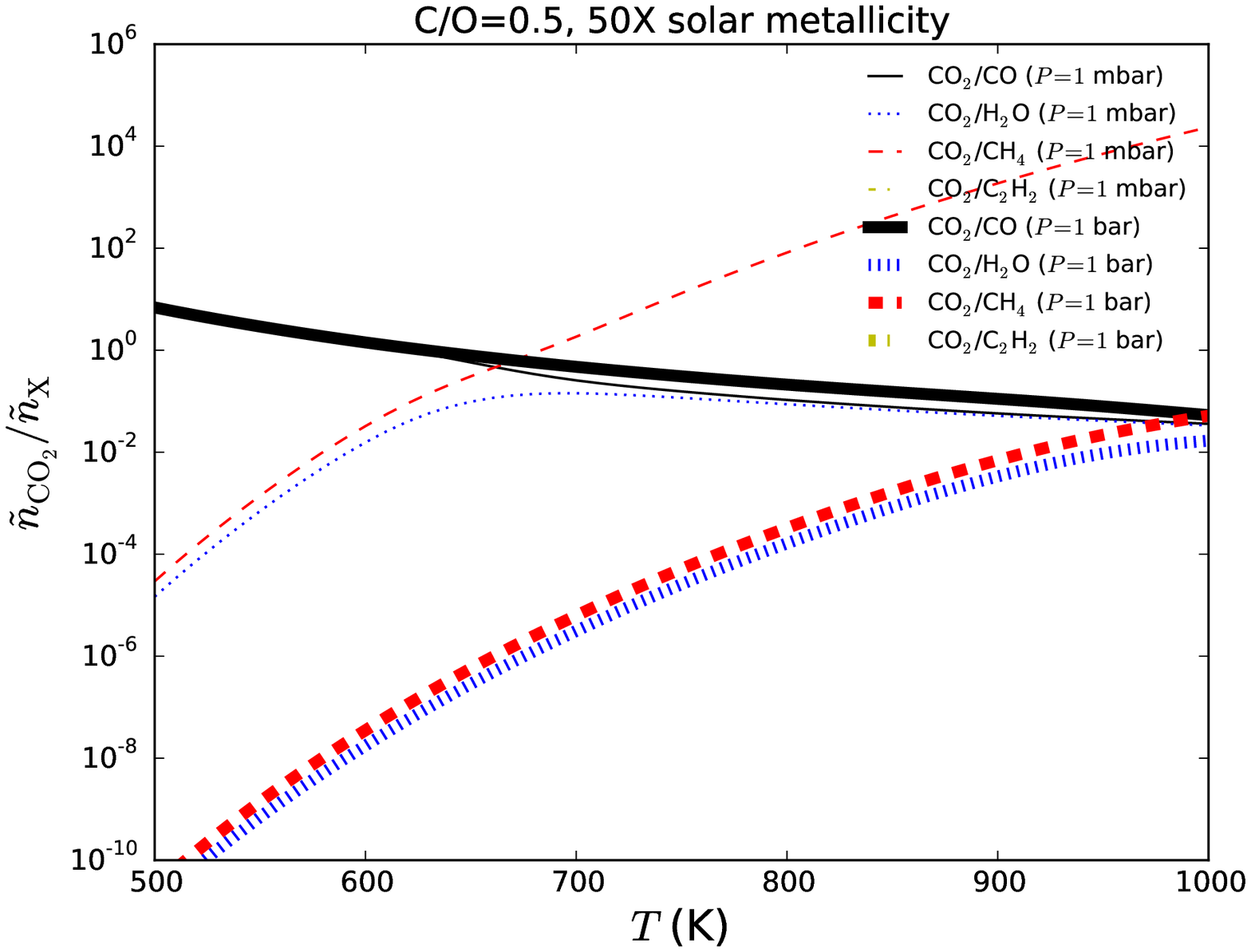}
\includegraphics[width=\columnwidth]{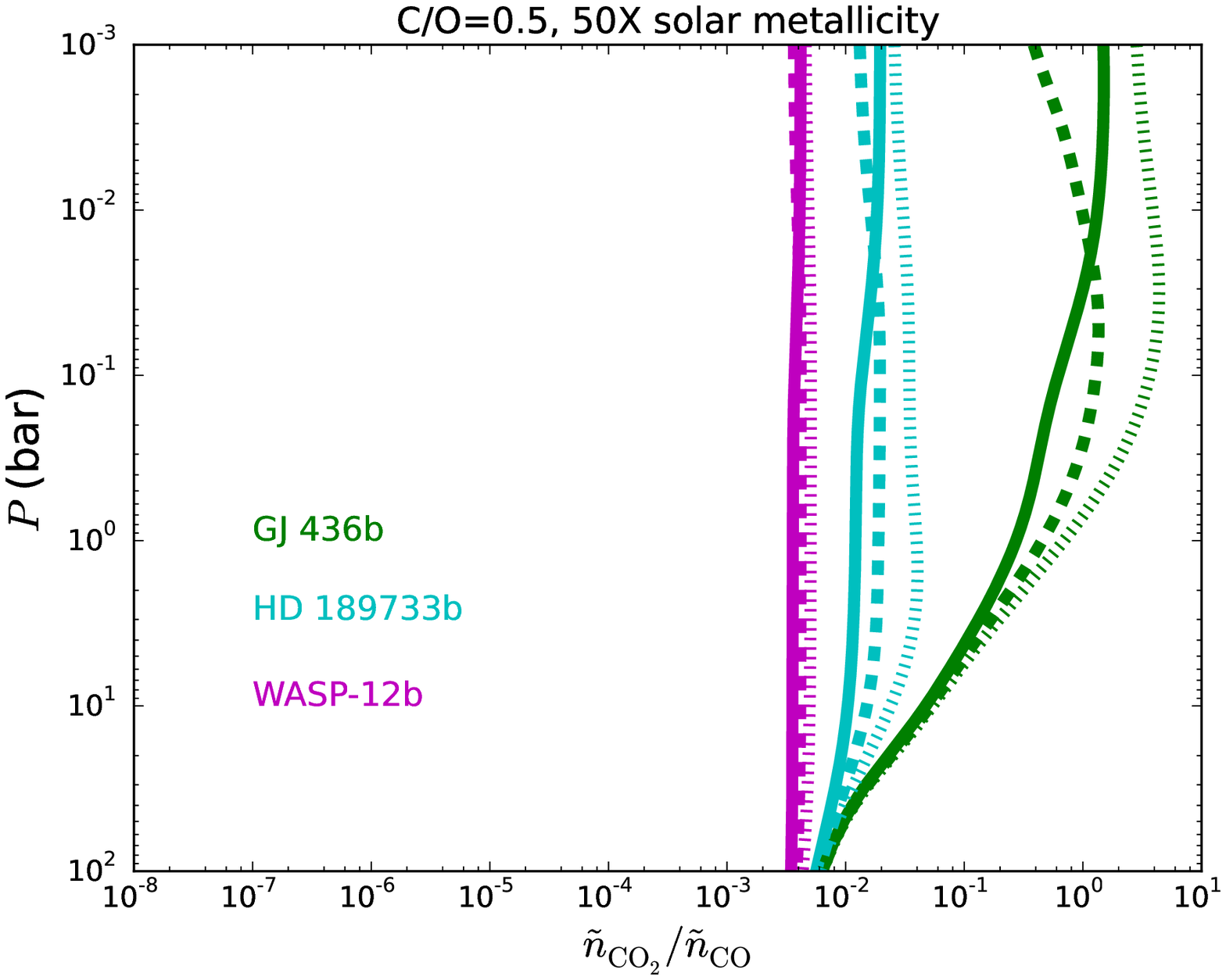}
\includegraphics[width=\columnwidth]{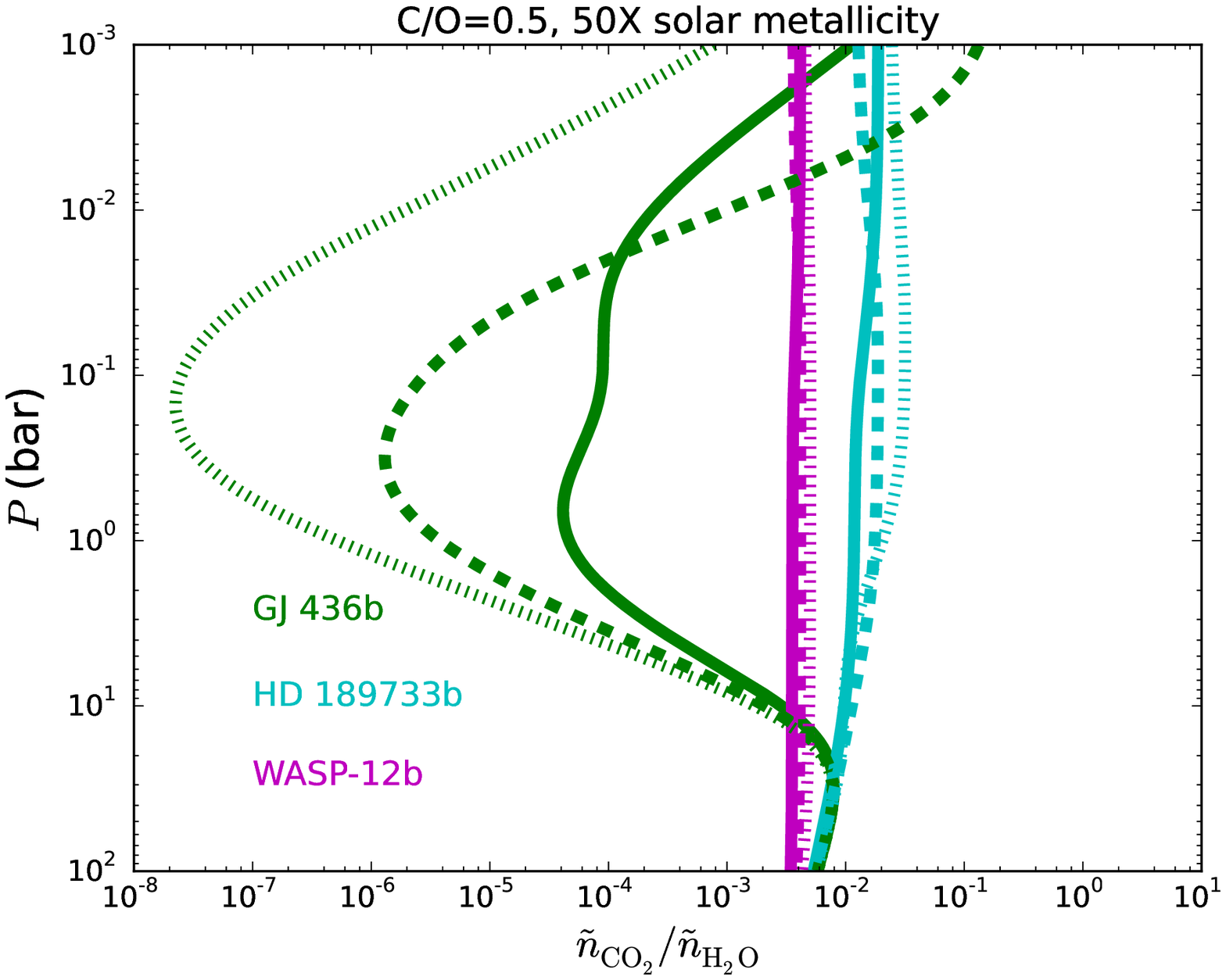}
\end{center}
%\vspace{-0.2in}
\caption{Relative molecular abundances (top-left panel) and their ratios (top-right panel) for enhanced metallicity ($\tilde{n}_{\rm O} = 2.5 \times 10^{-2}$) and $\mbox{C/O}=0.5$, meant to be representative of GJ 436b.  In the top-right panel, the ratio of carbon dioxide to acetylene exceeds the axis limit.  Also shown are the calculations in Figure \ref{fig:tp} repeated for $\tilde{n}_{\rm O} = 2.5 \times 10^{-2}$ and $\mbox{C/O}=0.5$ (bottom panels).}
%\vspace{0.3in}
\label{fig:gj436}
\end{figure*}

\subsection{Mixing ratios (molecular abundances normalized by that of molecular hydrogen)}

In Figure \ref{fig:abundances}, we show calculations of the mixing ratios of carbon dioxide, carbon monoxide, water, methane and acetylene.  We set $\tilde{n}_{\rm O}=5 \times 10^{-4}$, which is the approximate value at the solar photosphere.  For a sub-solar carbon abundance ($\mbox{C/O}=0.1$), water is the dominant molecule regardless of temperature \citep{moses13b}.  Methane dominates carbon monoxide at low temperatures \citep{bs99,lodders02} and the trend reverses at high temperatures depending on the pressure \citep{madhu12}.  Carbon dioxide is dominated by methane at low temperatures, but the trend reverses in a pressure-independent manner, because the number of reactants (two) versus products (two) is the same in the chemical reaction involving the production of carbon dioxide.  Acetylene always remains subdominant in a carbon-poor environment.  As we increase C/O to 1, 10 and 100, the system becomes water-poor \citep{madhu12}.  The trend between methane and carbon monoxide persists in a qualitative sense, but the transition occurs at different temperatures.  At high temperatures, acetylene starts to dominate both methane and carbon dioxide \citep{madhu12,moses13a,venot15}.

A key limitation of our approach is that graphite starts to condense out of the gas when the carbon-to-oxygen ratio and/or metallicity are greatly enhanced \citep{moses13b}.  For example, \cite{moses13b} have shown that, at $P=0.1$ bar, graphite formation occurs at $\mbox{C/O} \gtrsim 0.7$ and $T \gtrsim 700$ K when the metallicity is 300 times above solar; their Figure 7 also shows that, at solar $\mbox{C/O}$, graphite forms when the metallicity is 1000 times solar and $T \lesssim 800$ K.  Generally, graphite formation and rain out are expected to maintain the gas-phase $\mbox{C/O}$ near unity (see \citealt{moses13b} and references therein).  We note that \cite{hu14} also did not include graphite formation in their calculations.

\subsection{Ratios of mixing ratios}

In Figure \ref{fig:ratios}, we divide the mixing ratio of carbon dioxide by that of the other molecules.  The striking thing we notice is that carbon dioxide is always subdominant compared to carbon monoxide and water, across the entire range of temperatures considered, and for $\mbox{C/O}=0.1$, 1, 10 and 100.  Physically, the relative scarcity of carbon dioxide stems from the fact that it both requires the production of carbon monoxide and successfully competing with it \textit{and} water for oxygen atoms.  Furthermore, $\tilde{n}_{\rm CO_2}/\tilde{n}_{\rm CO}$ is insensitive to pressure for the reason previously mentioned, unless acetylene starts to become dominant.  By contrast, a higher pressure favors the product of methane over carbon dioxide as the reverse reaction is favored, because the number of products (four) exceeds the number of reactants (two)---this is a manifestation of Le Ch\^{a}telier's principle.  

\subsection{Chemistry in a broad range of model atmospheres}

Next, we wish to compute the abundance of carbon dioxide relative to carbon monoxide and water in model atmospheres and fully explore the effects of pressure variations.  We pick $\mbox{C/O}=0.5$ for illustration.  In Figure \ref{fig:tp}, we create temperature-pressure profiles of model atmospheres with three different equilibrium temperatures: $T_{\rm eq}=600$, 1200 and 2500 K.  These values are representative of a broad range of currently characterisable exoplanetary atmospheres from GJ 436b to HD 189733b to WASP-12b.  For our other parameters, we choose typical, plausible or illustrative values: $g=10^3$ cm s$^{-2}$, $T_{\rm int} = 300$ K, $\kappa_{\rm vis} = 0.01$ cm$^2$ g$^{-1}$ and $\kappa_{\rm IR} = 0.02$ cm$^2$ g$^{-1}$.  These choices imply that the infrared photosphere resides at a pressure of
\begin{equation}
P_{\rm IR} \sim \frac{g}{\kappa_{\rm IR}} = 50 \mbox{ mbar}.
\end{equation}
Starlight is mostly deposited at a pressure of $0.63 g / \kappa_{\rm vis}$ \citep{hhps12,hml14}.  For each value of $T_{\rm eq}$, we create three variations: pure absorption, a finite Bond albedo (which we set to $A_{\rm B}=0.5$) and a temperature inversion (which we create by doubling the value of the visible opacity to $\kappa_{\rm vis}=0.04$ cm$^2$ g$^{-1}$).  This gives us a total of 9 model atmospheres.

For each model atmosphere, we calculate $\tilde{n}_{\rm CO_2}/\tilde{n}_{\rm CO}$ and $\tilde{n}_{\rm CO_2}/\tilde{n}_{\rm H_2O}$.  Across all 9 models (see Figure \ref{fig:tp}), we find that the mixing ratio of carbon dioxide is always subdominant, compared to carbon monoxide and water, by at least an order of magnitude---often more.  Thus, we have demonstrated the conclusion previously stated in equation (\ref{eq:conclusion}), at least for elemental oxygen abundances that are solar.

\section{Discussion}

\subsection{Do atmospheric dynamics and mixing affect our conclusion?}

Our chemical-equilibrium calculations do not account for disequilibrium chemistry that may arise from atmospheric mixing.  Workers constructing one-dimensional models of the atmosphere have traditionally used a diffusion coefficient, commonly termed an ``eddy mixing coefficient", to collectively mimic advection, convection, diffusion and turbulence (e.g., \citealt{moses11,moses13a,moses13b,venot12,venot14,agundez14,hu14,hu15}).  In three dimensions, atmospheric circulation hardly resembles diffusion with equator-to-pole circulation cells that penetrate down to $\sim 10$ bar across the temperature range we are examining \citep{showman09,hmp11,hfp11,php12,kataria13,mayne14}.  In the case of GJ 436b, the atmospheric circulation penetrates down to only $\sim 1$ bar \citep{lewis10}.  (For a review of the atmospheric dynamics of hot exoplanetary atmospheres, see \citealt{hs15}.)  Furthermore, it is often overlooked that the intense stellar heating associated with close-in, transiting exoplanets flattens their temperature-pressure profiles and suppresses convection \citep{hfp11}.  Thus, the association between eddy diffusion/mixing and convection becomes even more tenuous.

Nevertheless, for the sake of discussion, we invoke the simplest approach for including atmospheric mixing: the ``quenching approximation", where one locates the point in the atmosphere where the chemical and dynamical timescales equate.  Above this point (i.e., at lower pressures or higher altitudes), the chemical abundances are frozen to their quench-point values.  A caveat is that different chemical species will generally have different quench points.  If carbon dioxide is subdominant compared to carbon monoxide and water \textit{everywhere} in the atmosphere, then locating the quench point becomes irrelevant---quenching will not alter this outcome.  Generally, the conclusion reached in equation (\ref{eq:conclusion}) is unaffected by atmospheric dynamics and mixing.  For atmospheric mixing to alter our conclusion requires substantially enhanced metallicities \textit{and} low temperatures---in \S\ref{subsect:metallicity}, we will quantify what ``substantially enhanced" and ``low" means.

\subsection{Does photochemistry strengthen or weaken our conclusion?}

Photochemistry is expected to enhance the production of acetylene \citep{line11,moses11},
\begin{equation}
\begin{split}
&2 \mbox{CO} + \mbox{H}_2 \rightarrow \mbox{C}_2\mbox{H}_2 + 2\mbox{O}, \\
&2 \mbox{CH}_4 \rightarrow \mbox{C}_2\mbox{H}_2 + 3\mbox{H}_2,
\end{split}
\end{equation}
where the first and second net reactions proceed via the photolysis of carbon monoxide (into C and O atoms, which act as radicals) and water (into the O atom and the OH radical), respectively.  The first reaction is relevant only for $P \lesssim 1$ $\mu$bar, while the second reaction is the dominant one for photospheric pressures \citep{moses11}.  The second net reaction is essentially the same as the one we consider in the present study for thermochemistry.  Thus, if it is enhanced via photochemistry, then it can only lead to enhanced acetylene production at the expense of carbon dioxide.  Overall, the presence of photochemistry strengthens, rather than weakens, our conclusion.

\subsection{The effects of enhanced metallicity}
\label{subsect:metallicity}

Our results, which are analytical except for the use of a numerical solver for the quintic equation, appear to be supported by more sophisticated calculations in the literature that focus on somewhat narrower regimes of parameter space (e.g., \citealt{moses11,koppa12,madhu12,venot12,venot14,hu14}).  Exceptions include \cite{line11}, who focused on the hot Neptune GJ 436b and studied the effects of enhanced metallicities; they reported $\tilde{n}_{\rm CO_2}/\tilde{n}_{\rm CO} \gtrsim 1$ at some locations in their model atmospheres with $50\times$ the solar metallicity.  \cite{moses13b} demonstrated that carbon dioxide could be the dominant gas by mass, over a temperature range consistent with that found in GJ 436b, if the elemental carbon abundance is enhanced by four orders of magnitude over its solar value.  \cite{agundez14} also noted that $\tilde{n}_{\rm CO_2}/\tilde{n}_{\rm H_2O}$ increases dramatically with metallicity.

We interpret an enhanced metallicity to be an increased value of the elemental abundance of oxygen ($\tilde{n}_{\rm O}$) relative to its solar value ($5 \times 10^{-4}$).  In Figure \ref{fig:gj436}, we repeat our calculations for an enhanced metallicity of $50\times$ solar, but we keep the carbon-to-oxygen ratio at its solar value ($\mbox{C/O}=0.5$) as metallicity and C/O are independent quantities.  We also focused on a lower temperature range that is representative of GJ 436b.  We are thus able to reproduce the same qualitative behavior as \cite{line11}: $\tilde{n}_{\rm CO_2}/\tilde{n}_{\rm CO} \gtrsim 1$ at $T \approx 500$--600 K.  The cross-over point appears to occur at about 600 K.  Despite the enhanced metallicity, our GJ 436b-like calculation displays $\tilde{n}_{\rm CO_2}/\tilde{n}_{\rm H_2O} \ll 1$ throughout.  We also repeated the calculations in Figure \ref{fig:tp} and find that $\tilde{n}_{\rm CO_2}/\tilde{n}_{\rm CO},\tilde{n}_{\rm CO_2}/\tilde{n}_{\rm H_2O} \ll 1$ for HD 189733b- and WASP-12b-like atmospheres, despite the enhanced metallicity.

\begin{figure}%[!h]
\begin{center}
%\vspace{-0.2in}
\includegraphics[width=\columnwidth]{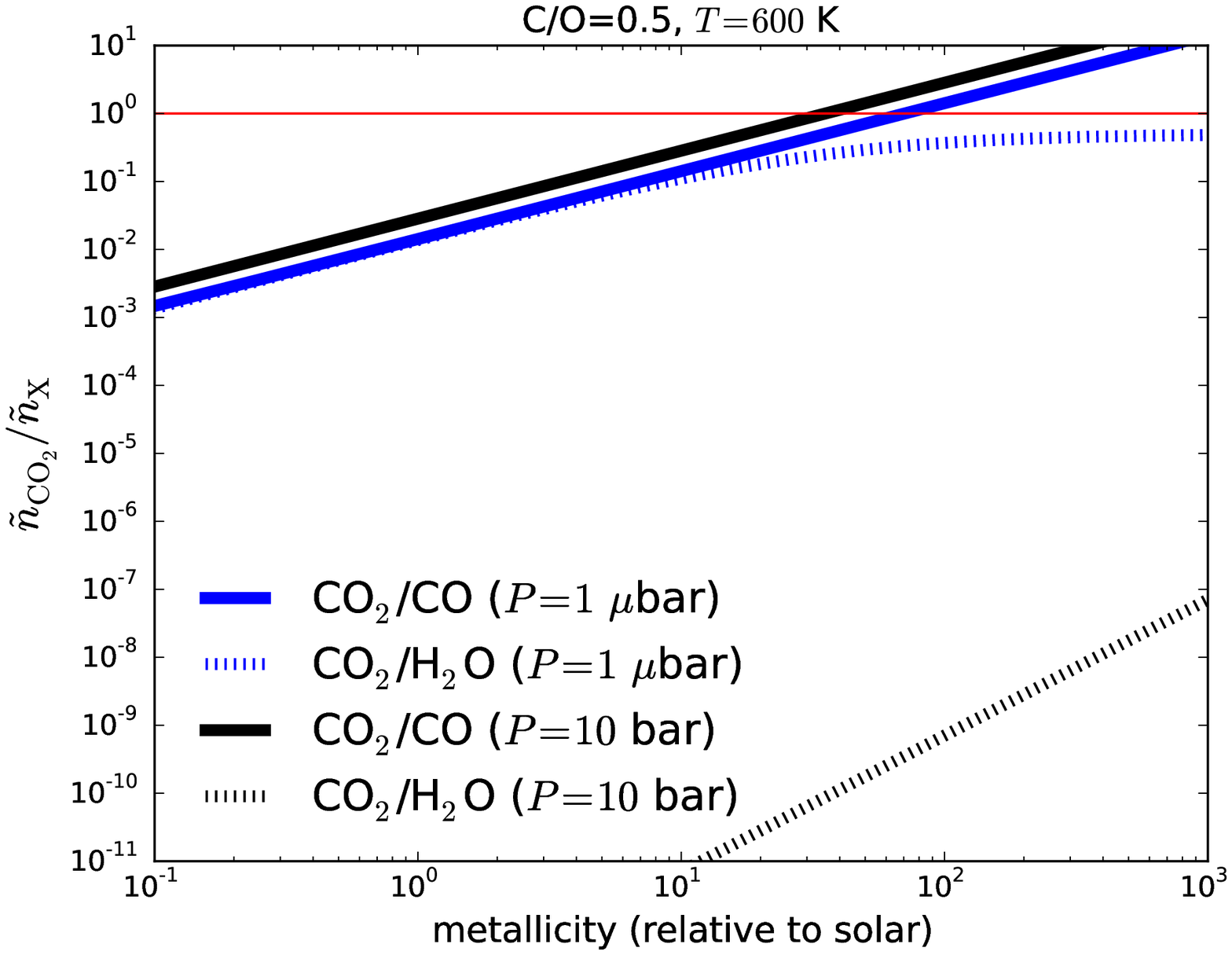}
\includegraphics[width=\columnwidth]{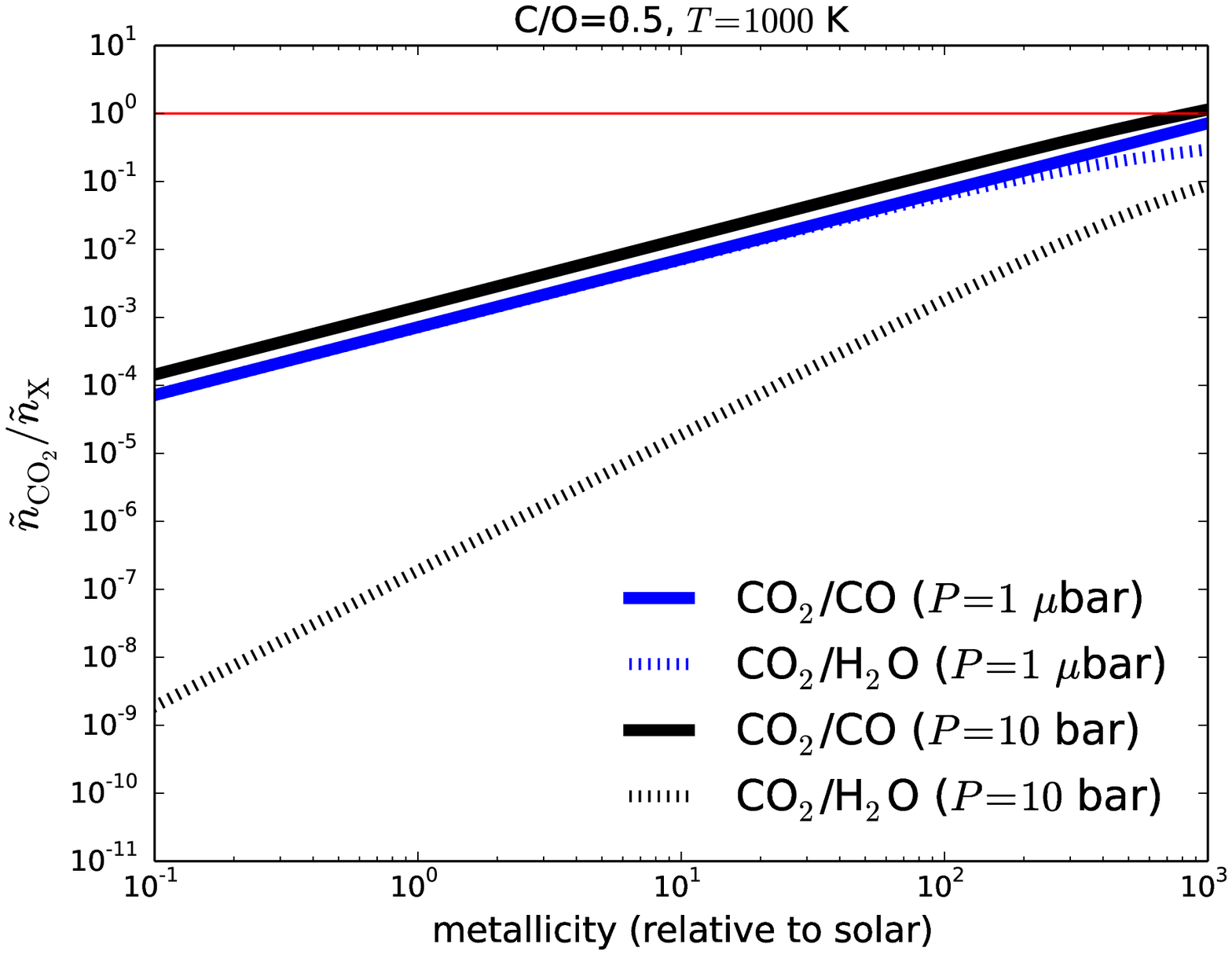}
\end{center}
%\vspace{-0.2in}
\caption{Mixing ratios of carbon dioxide relative to those of carbon monoxide and water for $T=600$ K (top panel) and 1000 K (bottom panel) and the full range of pressures that is relevant to infrared photospheric emission and atmospheric circulation in exoplanetary atmospheres ($1 ~\mu\mbox{bar} \le P \le 10$ bar).  We assume a solar carbon-to-oxygen ratio ($\mbox{C/O}=0.5$).}
%\vspace{0.1in}
\label{fig:transition}
\end{figure}

Figure \ref{fig:transition} expresses the effects of metallicity in a more general way.  We examine pressures of 1 $\mu$bar and 10 bar, because these values bracket the conceivable range of pressures probed by the infrared observations.  We examine pressures up to 10 bar, because this is the maximum depth to which atmospheric circulation penetrates for the range of equilibrium temperatures we are interested in \citep{php12}.  We see that atmospheres that have $T=600$ K always have $\tilde{n}_{\rm CO_2}/\tilde{n}_{\rm H_2O} < 1$ and only have $\tilde{n}_{\rm CO_2}/\tilde{n}_{\rm CO} > 1$ when the metallicity is enhanced by more than an order of magnitude relative to solar.  By the time the temperature reaches $T=1000$ K, we rarely have $\tilde{n}_{\rm CO_2}/\tilde{n}_{\rm CO} > 1$ unless one entertains implausible metallicities reaching $\sim 10^3$ times that of solar.

A few considerations give pause to contemplating such high metallicities.  To begin with, we note that the elemental abundances of carbon and oxygen in Jupiter are enhanced by only about an order of magnitude relative to solar \citep{wong04}.  \cite{mf11} have shown, using interior structure calculations applied to mass and radius measurements, that gas-giant exoplanets have bulk enhancements of metallicities $\sim 1$--10 times of their parent stars.  Furthermore, carbon-rich stars appear to be rare \citep{teske14}.

\subsection{Consequences and implications for retrieval studies}

Our results have important consequences for retrieval studies, which is that chemical abundances cannot be allowed to completely roam free.  For atmospheres with temperatures of 1000 K and above, one may wish to set, as a prior in one's retrieval procedure, that the mixing ratio of CO$_2$ cannot exceed that of CO and H$_2$O,
\begin{equation}
\frac{\tilde{n}_{\rm CO_2}}{\tilde{n}_{\rm CO}} < 1 \mbox{ and } \frac{\tilde{n}_{\rm CO_2}}{\tilde{n}_{\rm H_2O}} < 1.
\label{eq:priors}
\end{equation}
Retrieval outcomes that violate these criteria, for hot atmospheres, should be flagged as being unphysical.  

We join the debate on the atmosphere of the hot Jupiter WASP-12b.  \cite{madhu11} has previously reported that $\mbox{C/O}>1$.  Their Figure 1 shows four best-fit models with $\tilde{n}_{\rm CO_2}/\tilde{n}_{\rm CO},\tilde{n}_{\rm CO_2}/\tilde{n}_{\rm H_2O} \ll 1$, but an inspection of their Figure 2 reveals that, while $\tilde{n}_{\rm CO_2}/\tilde{n}_{\rm CO}<1$ appears to hold for their ensemble of best-fit models (which are colored purple), the same models also seem to have $\tilde{n}_{\rm CO_2}/\tilde{n}_{\rm H_2O}>1$.  By contrast, \cite{line14} reported that $\mbox{C/O}<1$ for WASP-12b, but this appears to be based on retrievals that yield $\tilde{n}_{\rm CO_2}/\tilde{n}_{\rm CO},\tilde{n}_{\rm CO_2}/\tilde{n}_{\rm H_2O} \gg 1$, as stated in their Table 3 and displayed in their Figure 3.  Furthermore, \cite{stevenson14} have claimed that \cite{line14} do not include acetylene or hydrogen cyanide in their analysis, which drives the retrieval calculations to ``unrealistically-large" abundances of CO$_2$.  It is thus unsurprising that the two sets of authors reach different conclusions regarding the carbon-to-oxygen ratio of WASP-12b, even without discussing the controversy surrounding the data themselves \citep{crossfield12,stevenson14}.  In other words, some of these reported retrievals are chemically impossible (but see \citealt{moses13a}).  To be fair, \cite{line14} did remark: ``One might question how realistic solutions with such high abundances of CO$_2$ may be. Such high CO$_2$ abundances are generally not thermochemically permissible in highly reducing hot-Jupiter-like atmospheres."  \cite{line14} then performed a separate retrieval analysis by setting a hard limit on the mixing ratio of CO$_2$ to be $10^{-5}$ and found that it has no effect on the CH$_4$ and CO abundances, although the retrieved H$_2$O abundance is increased by 2 to 3 orders of magnitude.

\cite{stevenson14} present an improved analysis of the dayside emission of WASP-12b, which included new data from the Spitzer Space Telescope.  They did not enforce the priors in equation (\ref{eq:priors}), so as to allow for uncertainties associated with the absorption opacities and for the possibility of enhanced helium abundances.  This permits both physically plausible and implausible retrievals to be reported and for the rejection of implausible solutions after the fact.  \cite{stevenson14} concluded that oxygen-rich atmospheres require implausibly-high abundances of carbon dioxide.  Both \cite{madhu12} and \cite{stevenson14} concluded that the dayside emission spectrum of WASP-12b favors a carbon-rich interpretation.

More recently, \cite{kreidberg15} obtained a near-infrared transmission spectrum, which probes a series of water lines originating from the terminators of the atmosphere of WASP-12b, using the Wide Field Camera 3 on the Hubble Space Telescope.  Using a thermochemically-consistent retrieval analysis, they ruled out a carbon-rich scenario for the terminator region of WASP-12b at more than $3\sigma$-level confidence.  

Within the context of our study, we find that the atmosphere of WASP-12b is too hot for enhanced metallicity to play a role (see Figures \ref{fig:tp}, \ref{fig:gj436} and \ref{fig:transition}) in boosting the CO$_2$ abundance over that of CO and H$_2$O.  If we repeat our calculations in Figure \ref{fig:transition} for $T=2000$--3000 K (not shown), we obtain the result that the metallicity needs to be enhanced by about 4 orders of magnitude for WASP-12b---a statement that is independent of pressure, because of the high temperatures involved.  Another consequence is that, since $\tilde{n}_{\rm CO_2}/\tilde{n}_{\rm C_2H_2}$ appears to vary from $\gg 1$ to $\ll 1$ as C/O increases, this pair of molecules may be used as a diagnostic for the carbon-to-oxygen ratio (see also \citealt{venot15}), although this will be challenging as their spectral lines will probably be dominated by those of the other molecules.

Generally, past claims in the literature of $\tilde{n}_{\rm CO_2}/\tilde{n}_{\rm CO} \gg 1$ and/or $\tilde{n}_{\rm CO_2}/\tilde{n}_{\rm H_2O} \gg 1$ should be viewed with skepticism, unless a chemical mechanism (e.g., enhanced metallicity at low temperatures) has been elucidated.  If CO$_2$ is reported to be constrained by a retrieval calculation but CO is not, then it should also be viewed skeptically.

\acknowledgments
The motivation for this study originated from a discussion between KH and Nikku Madhusudhan during KH's visit to Cambridge University in December 2014.  We thank Mike Line, Julie Moses and Olivia Venot for constructive and helpful comments on an earlier version of this manuscript, as well as the two anonymous referees for constructive and gracious reports.  KH acknowledges financial and secretarial support from the Center for Space and Habitability, the PlanetS NCCR (National Center of Competence in Research) framework, the Universities of Bern and Zurich and the Swiss-based MERAC Foundation.  KH is grateful to Claudia for encouragement and support.

%%% REFERENCES %%%

\appendix

\section{Notes on Global versus Local Energy Conservation in One-dimensional Atmospheres}

In the limit where advection, ``$PdV$ work" and thermal conduction may be neglected, the first law of thermodynamics, which is a statement of energy conservation, states that \citep{hml14}
\begin{equation}
\frac{\partial T}{\partial t} = \frac{1}{c_P} \frac{\partial {\cal F}_-}{\partial \tilde{m}},
\label{eq:first_law}
\end{equation}
where $T$ is the temperature, $t$ denotes the time, $c_P$ is the specific heat capacity at constant pressure and $\tilde{m}$ is the column mass.  The quantity ${\cal F}_-$ is the net flux, which is the difference in flux entering and exiting each layer of the model atmosphere, integrated over all wavelengths.  In a hydrostatic atmosphere, the column mass and pressure ($P$) are straightforwardly related by $P= \tilde{m} g$, where $g$ is the surface gravity.

Since opacities are temperature-dependent and temperatures depend on the opacities, one needs to iterate between them in a model atmosphere.  With each iteration, one may compute the fluxes in each layer and thus the net flux.  By using equation (\ref{eq:first_law}), one may then update the temperature-pressure profile during each iteration.  The model atmosphere reaches a steady state, where the temperature in each layer ceases to change, when the gradient of the net flux is zero.  This is known as radiative equilibrium.  It is a statement of \textit{local} energy conservation, since it applies to each layer of the atmosphere.

Global energy conservation involves considering the energy budget of the atmosphere and ensuring that the total amount of energy entering and exiting it must be conserved.  By taking the first moment of the radiative transfer equation, one may show that the energy content, per unit mass and time and integrated over all wavelengths, is \citep{hml14}
\begin{equation}
Q = \frac{\partial {\cal F}_-}{\partial \tilde{m}} = \kappa_{\rm J} {\cal J} - 4 \kappa_{\rm P} \sigma_{\rm SB} T^4,
\label{eq:energy_conserve}
\end{equation}
where $\kappa_{\rm J}$ is the absorption mean opacity, $\kappa_{\rm P}$ is the Planck mean opacity, ${\cal J}$ is the total intensity integrated over all wavelengths and $\sigma_{\rm SB}$ is the Stefan-Boltzmann constant.  Equation (\ref{eq:energy_conserve}) still applies in a layer-wise fashion, with the first term (after the second equality) accounting for scattered light and also flux entering and exiting the layer from adjacent layers.  The second term accounts for the thermal emission of each layer.  Within the context of the method of moments, we emphasize that no approximations have been taken on equation (\ref{eq:energy_conserve}).

To account for \textit{global} energy conservation, we need to integrate the model atmosphere over column mass,
\begin{equation}
\int^\infty_{\tilde{m}} Q ~d\tilde{m} = \tilde{Q}\left(\tilde{m},\infty\right),
\label{eq:energy_conserve_1b}
\end{equation}
where we have defined, for any arbitrary pair of column masses $\tilde{m}_1$ and $\tilde{m}_2$,
\begin{equation}
\tilde{Q}\left(\tilde{m}_1,\tilde{m}_2\right) \equiv \int^{\tilde{m}_2}_{\tilde{m}_1} Q ~d\tilde{m}.
\end{equation}

We now have the mathematical machinery to describe local versus global energy conservation.  Local energy conservation or radiative equilibrium occurs when $Q=0$.  This in turn implies that $\tilde{Q}(0,\infty)=0$, as shown by equation (\ref{eq:energy_conserve_1b}).  But enforcing global energy conservation alone ($\tilde{Q}(0,\infty)=0$) does not guarantee radiative equilibrium ($Q=0$).  Mathematically, the gradient of the wavelength-integrated net flux may conspire to be non-zero at multiple locations, only to cancel out when summed over the entire atmosphere, such that $\tilde{Q}(0,\infty)=0$ but $Q \ne 0$.

In other words, radiative equilibrium is a necessary and sufficient condition for global energy conservation.  However, global energy conservation is an insufficient condition for local energy conservation.  

\label{lastpage}

\end{document}